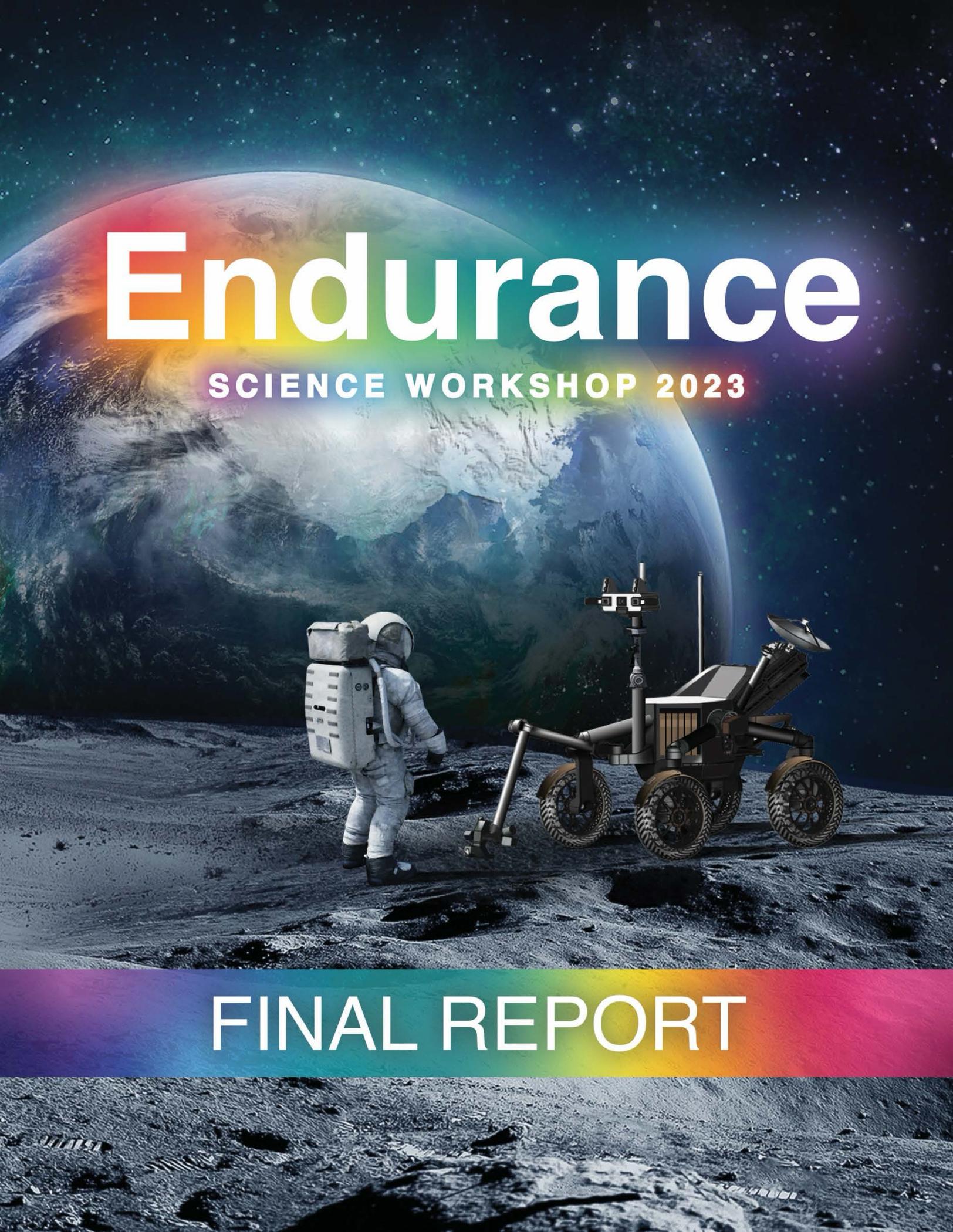

# Endurance
## SCIENCE WORKSHOP 2023

**FINAL REPORT**

# FINAL REPORT OF THE ENDURANCE SCIENCE WORKSHOP 2023

**SCIENCE ORGANIZING COMMITTEE:**

James Tuttle Keane* (Jet Propulsion Laboratory, California Institute of Technology)

Barbara Cohen (NASA Goddard Space Flight Center)

Carolyn Crow (University of Colorado Boulder)

Benjamin Greenhagen (Johns Hopkins University, Applied Physics Laboratory)

Bradley Jolliff (Washington University in St. Louis)

Yang Liu (Jet Propulsion Laboratory, California Institute of Technology)

Charles "Chip" Shearer (University of New Mexico)

Sonia Tikoo (Stanford University)

Sarah Valencia (NASA Goddard Space Flight Center, University of Maryland College Park)

**\*CORRESPONDING AUTHOR:** james.t.keane@jpl.nasa.gov.

**WEBSITE:** https://www.hou.usra.edu/meetings/endurance2023/.

**BACK COVER IMAGE:** Artist concept of the Endurance rover. Figure adapted from the Endurance concept study report. Artwork by David Hinkle, modified by James Tuttle Keane.

**FRONT COVER IMAGE:** Image of the lunar farside, including the northern portion of the South Pole–Aitken basin (bottom left quadrant of the visible face of the Moon), acquired by NASA's Artemis 1 Orion spacecraft on 21 November 2022.

# TABLE OF CONTENTS

# 1. EXECUTIVE SUMMARY

Endurance is a mission concept to explore and ultimately return samples from the Moon's largest and oldest impact basin, South Pole–Aitken (SPA). SPA holds the answers to many outstanding planetary science questions, including the earliest impact bombardment and dynamics of the Solar System and the evolution of the interior structure of the Moon. Endurance would address these questions by traversing 2,000 kilometers across the lunar farside, collecting a large mass of samples (up to 100 kilograms), and delivering those samples to Artemis astronauts for return to Earth. Endurance was identified as the highest priority strategic mission for NASA's Lunar Discovery and Exploration Program in the recent Planetary Science and Astrobiology Decadal Survey 2023–2032.

This report summarizes a three-day hybrid workshop (9–11 August 2023) about the science motivating the Endurance concept. This workshop was the first dedicated workshop to bring together scientists and engineers to discuss this specific mission concept. Topics discussed included next-generation sample analysis, instruments, traverses, in situ science, technical challenges, synergies with human exploration, and more. Over 160 individuals participated in the Endurance Science Workshop.

The Endurance Science Workshop organizing committee identified six major findings, which are detailed in this report. They are (in no priority order):

**Finding #1** Endurance is an exciting concept that would address long-standing, high-priority lunar and planetary science questions—and the community is ready for it. (Page 23–26)

**Finding #2** Endurance's sample science objectives are achievable, although they would require coordinated analysis techniques and numerous diverse samples. (Page 27–30)

**Finding #3** Geologic context is essential for addressing Endurance's science objectives. (Page 31–33)

**Finding #4** While Endurance's objectives center on sample return, Endurance's long traverse would enable a variety of additional transformative science investigations. (Page 34–41)

**Finding #5** Endurance is an ambitious mission that would be enabled and enhanced by investing in developing key technologies now. (Page 42–50)

**Finding #6** Endurance should strive to include more diverse perspectives in its formulation—particularly from early-career scientists and engineers who will ultimately operate the rover and analyze the samples. (Page 51–52)

Endurance is early in its formulation (pre-phase A) and the next major activity will be a Science Definition Team (SDT). An SDT is important for strategic missions as it defines the science requirements and science traceability. It is expected that this report and the findings therein may be useful input to the Endurance SDT.

# 2 THE ENDURANCE CONCEPT

Endurance is a mission concept for a long-range lunar rover to explore the Moon's largest, deepest, and oldest impact basin—South Pole–Aitken (SPA)—collect a large mass of samples from diverse terrains and deliver those samples to Artemis astronauts for return to Earth. This concept was developed during the recent Decadal Survey, *Origins, Worlds, and Life: A Decadal Strategy for Planetary Science and Astrobiology* 2023–2032 [1] (henceforth, "the Decadal Survey"), which ultimately recommended Endurance for implementation as a medium-sized (New Frontiers-class) strategic mission, and the highest priority for NASA's Lunar Discovery and Exploration Program (LDEP).

In this section, we provide a summary of the Endurance concept, including the scientific rationale, current implementation concept, and programmatic context. This information is important context for the findings and recommendations of the Workshop (**Section 4**). For more detail about the scientific and technical aspects of the Endurance concept, we refer the reader to the full Endurance concept study report[2], which was publicly released alongside the Decadal Survey.

## 2.1 WHY GO TO THE SOUTH POLE–AITKEN (SPA) BASIN?

South Pole–Aitken (SPA) is the largest, deepest, and oldest (undisputed) impact basin on the Moon. SPA spans ~2,000 kilometers across the southern lunar farside—from nearly the equator to the South Pole, with a depth of ~8 kilometers (**Figures 1**–**3**).

As the oldest preserved basin on the Moon, SPA is a critical datum constraining the impact bombardment record of the early Solar System. Determining the age of SPA, and the other large farside impact basins superposing it (*e.g.*, Poincaré, Schrödinger, Apollo) would provide important new constraints on impact bombardment of the Earth-Moon system during the time when life first emerged on Earth, and enable tests of long-standing, widely debated hypotheses about the so-called "Late Heavy Bombardment" or "Terminal Lunar Cataclysm". The lunar cataclysm hypothesis proposes that there was a spike of large impacts on the Earth and Moon around 3.9 billion years ago, linked to the orbital migration (and destabilization) of the giant planets. Thus, SPA is of Solar System-wide importance.

As the largest and deepest basin on the Moon, SPA likely excavated deep into the lower crust and mantle of the Moon—providing a window into the lunar interior and the overall thermochemical evolution of rocky worlds. At present, there is substantial debate about the composition and structure of the lunar mantle and whether/where this material is exposed at the surface. Samples of mantle (or mantle-derived) material would address this debate and test competing hypotheses about the thermochemical evolution of the Moon. The Moon is often viewed as an endmember for rocky planet thermochemical evolution, with the "lunar magma ocean" hypothesis defining the paradigm for over half a century. Testing these hypotheses would thus have implications for many rocky worlds. Finally, by virtue of being on the lunar farside, exploration of SPA would shed light on the three-dimensional aspects of thermochemical evolution of the Moon. The Moon exhibits a nearside-farside asymmetry of heat-producing elements, volcanism, and crustal structure that remains unexplained. Samples with known provenance from the farside and in situ measurements may be able to distinguish between competing hypotheses for the origin of this asymmetry.

---

[1] National Academies of Science, Engineering, and Medicine: Origins, Worlds, and Life: A Decadal Strategy for Planetary Science and Astrobiology 2023–2032 (2022): https://www.nationalacademies.org/our-work/planetary-science-and-astrobiology-decadal-survey-2023-2032. This document is referred to as "the Decadal Survey" in this report.

[2] Endurance concept study report (and all other Decadal Survey concept study reports): https://tinyurl.com/2p88fx4f and https://science.nasa.gov/planetary-science/resources/documents/.

Because of the scientific importance of SPA, SPA sample return has been the highest priority lunar activity for the last three Planetary Science (and Astrobiology) Decadal Surveys, and numerous other community and strategic documents. SPA sample return concepts have been proposed to competed mission opportunities (*e.g.*, NASA's New Frontiers program) multiple times in the past, although never selected for flight.

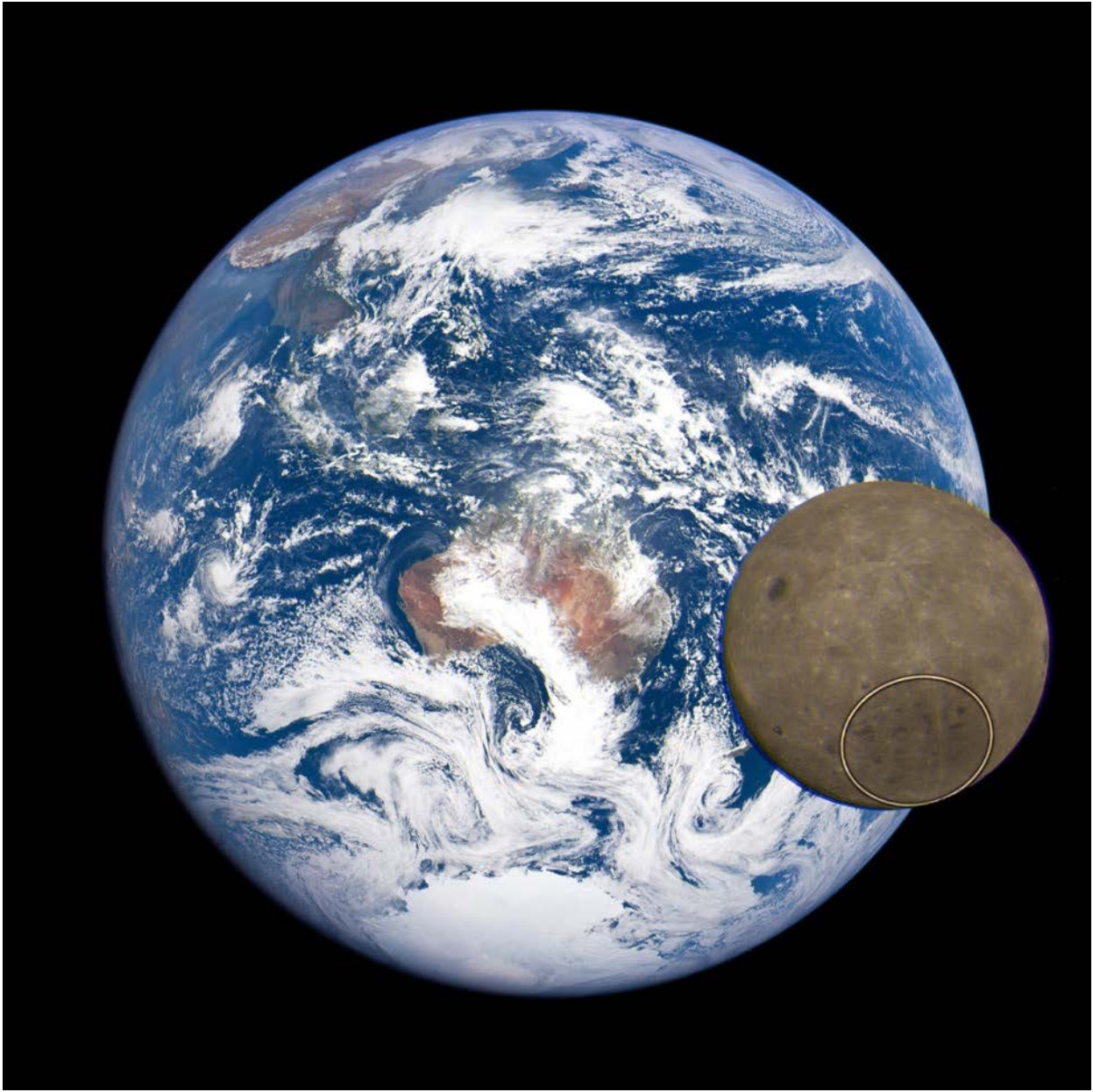

*Figure 1: The South Pole–Aitken (SPA) basin, as seen from space.* Deep Space Climate Observatory (DSCOVR) Earth Polychromatic Imaging Camera (EPIC) image of the Moon transiting the Earth on February 11th, 2021 (NASA, NOAA). The white circle encloses the approximate location of SPA.

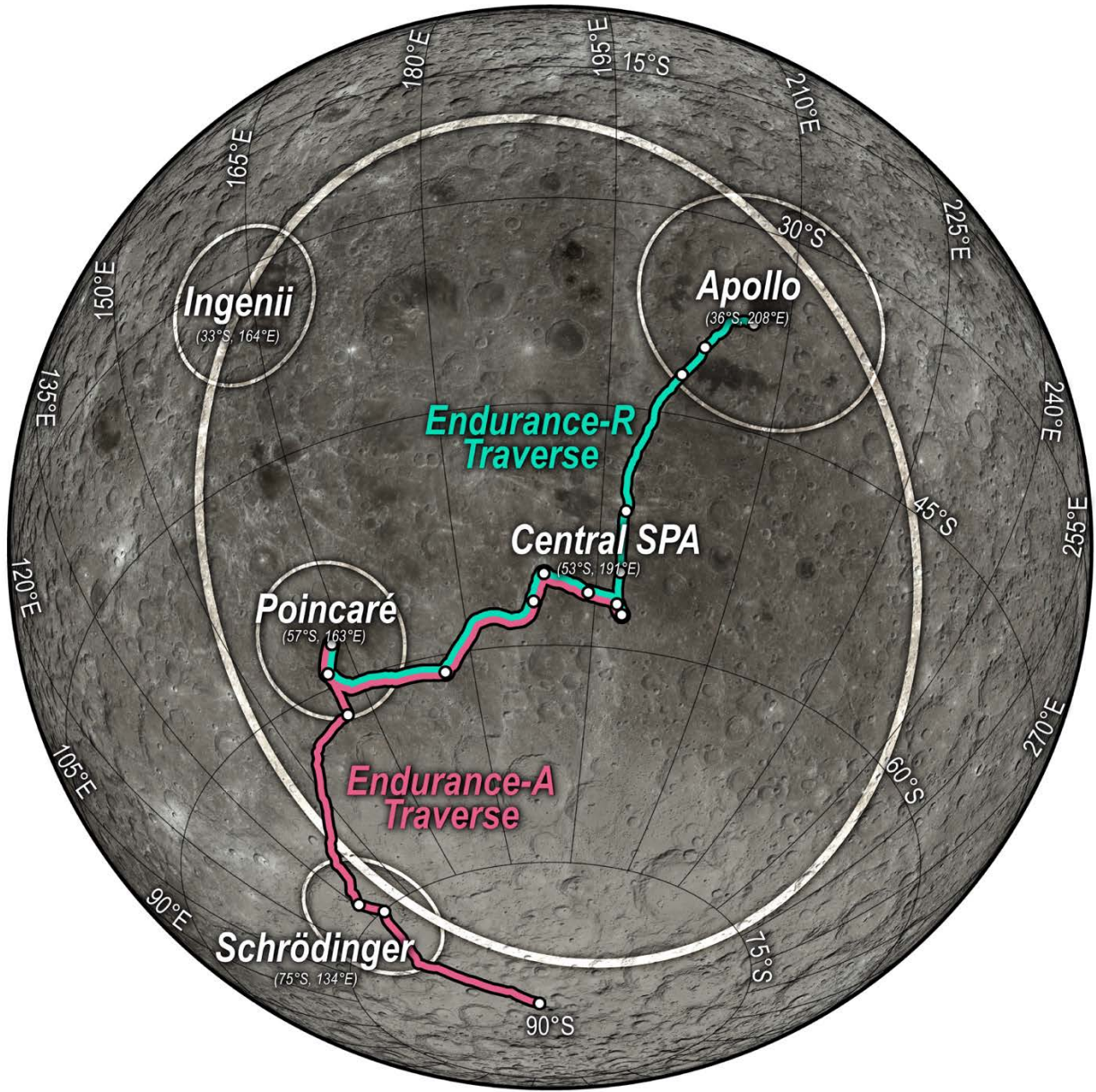

*Figure 2: Color image mosaic map of the South Pole–Aitken (SPA) basin and candidate Endurance rover traverses.* Perspective view of the Moon centered on SPA (54°S, 190°E) from an altitude of one lunar radius. The basemap is a color image mosaic with shaded relief (LROC, LOLA, NASA's Scientific Visualization Studio). The pink and green lines indicate the two candidate traverses for Endurance-A and Endurance-R, respectively. These two variants are described in **Section 2.2**. Endurance-A is the current baseline. Small white dots indicate potential sample sites. The larger white ellipses trace notable impact basins—with the largest outlining SPA (as mapped by Garrick-Bethell & Zuber 2009). Figure adapted from the Endurance concept study report[2].

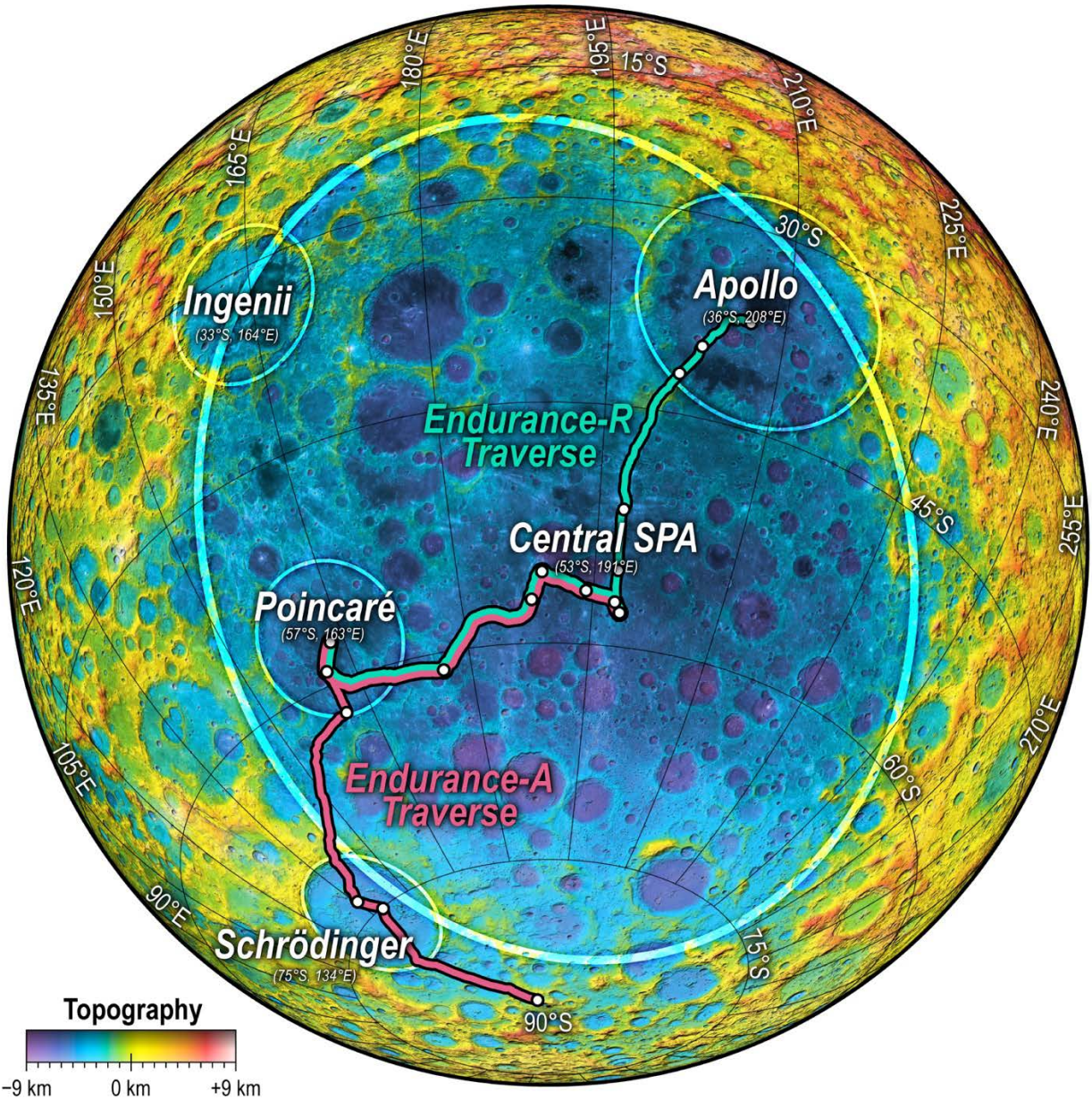

*Figure 3: Topographic map of the South Pole–Aitken (SPA) basin and candidate Endurance rover traverses. Perspective view of the Moon centered on SPA (54°S, 190°E) from an altitude of one lunar radius. The basemap shows a topographic view (LOLA, Selene). The pink and green lines indicate the two candidate traverses for Endurance-A and Endurance-R, respectively. These two variants are described in* **Section 2.2**. *Endurance-A is the current baseline. Small white dots indicate potential sample sites. The larger white ellipses trace notable impact basins—with the largest outlining SPA (as mapped by Garrick-Bethell & Zuber 2009). Figure adapted from the Endurance concept study report[2].*

## 2.2 THE ENDURANCE ROVER CONCEPT AS STUDIED BY THE DECADAL SURVEY

The Endurance concept originated in the *Origins, Worlds, and Life* Decadal Survey[1], and was designed to address the long-standing, high-priority science objectives that require sample return from SPA (**Table 1**).

| ENDURANCE SCIENCE OBJECTIVES | |
|---|---|
| 1 | Determine the age of the largest and oldest impact basin on the Moon, South Pole–Aitken (SPA) to anchor the earliest impact history of the Solar System. |
| 2 | Determine when post-SPA farside basins formed to test the giant planet migration and terminal cataclysm hypotheses, and to better constrain the inner solar system impact chronology used to date the surfaces of other planetary bodies. |
| 3 | Determine the age and mineralogical and geochemical composition of deep and crustal materials exposed in SPA to understand the bulk composition of the Moon, its primordial differentiation and geologic evolution, and the significance of chronologic measurements completed on nearside samples for timing lunar solidification. |
| 4 | Determine the age and nature of volcanic features and compositional anomalies on the lunar farside to characterize the thermochemical evolution of terrestrial worlds and constrain the origin of the Moon's nearside-farside asymmetry. |
| 5 | Determine the geologic diversity of the SPA terrane to provide geologic context for returned samples, ground truth for orbital measurements, and characterize the surface processes that shape planetary bodies. |

*Table 1: Endurance science objectives, as defined by the Decadal Survey[1] (page 574). Note that a future SDT (**Box 1**) may redefine these objectives.*

Endurance follows on a long history of mission concepts to return samples from SPA, but with a novel approach. Most past concepts employed a single lander at one location, acquiring samples at that site, and return the samples directly to Earth with a robotic return vehicle (akin to past robotic sample return missions: Luna 16, Luna 20, Luna 24, Chang'e 5, Chang'e 6). Instead, Endurance would utilize a long-range rover (**Figures 4–5**) landed on a commercial lander, traverse nearly 2,000 kilometers of lunar terrain (**Figures 2–3**), collect a large mass of samples (up to 100 kilograms) from 12 sites, and deliver those samples to Artemis astronauts at the lunar South Pole for return to Earth (**Figure 5**). Long-range mobility and a large sample cache (comparable to the mass returned from many Apollo missions) mitigates many of the concerns associated with past mission concepts. In the words of the Decadal Survey:

> *"[A]chieving the top science objectives with a fixed lander, as has been typically envisioned, is challenging. The committee concluded that the Endurance-A rover mission is a superior approach for acquiring abundant samples across diverse terrains to address multiple top-level science questions for the Moon and the solar system."* (Decadal Survey, page 592)

Endurance is an evolution of the Intrepid pre-Decadal Planetary Mission Concept Study[3], which conceived of a long-range rover for exploring the nearside Procellarum region to investigate the Moon's magmatic history. Intrepid was paradigm-changing, as it established that long-range rovers were plausible.

The baseline Endurance concept—as designed during the Decadal Survey concept study (**Figures 4–5**)—would consist of a large (mass: ~500 kilograms, height: 2.5 meters) rover that would be delivered to the lunar surface by a commercial lander via NASA's Commercial Lunar Payload Service (CLPS)[4]. CLPS landers with this capability are in development for the VIPER rover mission[5], Artemis, and other future missions. Endurance would have a robust mobility system, capable of driving in day and night, with 4-wheeled driving and steering, a top speed of 1 kilometer per hour, and traversing slopes up to 20° (more than sufficient for traversing SPA). The traverse would be accomplished within 4 years, including over 1 year in margin. The baseline Endurance concept would use a NextGen Mod 1 Radioisotope Thermoelectric Generator (RTG), with secondary battery, to power the traverse. Communications would utilize an orbital farside communications relay. This relay was not designed in the Endurance concept study, but multiple such relays are anticipated to be operational during the timeframe of the mission (early 2030s). Endurance would collect its samples using a large scoop, storing samples from each of the 12 sample sites in separate, detachable sample containers. Endurance would fly a suite of instruments to both provide geologic context, and aid in sample selection. An important aspect of the Endurance concept study is that no instrument trade was performed; rather, it was assumed that Endurance would use the Intrepid instrument

---

[3] Intrepid Planetary Mission Concept Study Report (2020): https://science.nasa.gov/science-red/s3fs-public/atoms/files/Lunar%20INTREPID.pdf.

[4] Commercial Lunar Payload Services (CLPS): https://www.nasa.gov/commercial-lunar-payload-services/.

[5] Volatiles Investigating Polar Exploration Rover (VIPER): https://science.nasa.gov/mission/viper/. Note that at the time of this writing this report, NASA has cancelled VIPER (https://www.nasa.gov/news-release/nasa-ends-viper-project-continues-moon-exploration/) and its future is uncertain. Nonetheless, Astrobotic—the company tasked with landing VIPER—is anticipated to still demonstrate its Griffin lander in the near future.

suite—including color cameras, infrared spectrometer, hand lens imager, gamma-ray and x-ray spectrometers, and particles and fields instruments—as detailed in **Table 2**. This decision (stipulated by the Decadal Survey) was made on the assumption that the Intrepid instrument suite would likely be sufficient for addressing all of Endurance's science objectives. A future SDT (**Box 1**) is necessary to define Endurance's measurement requirements and a notional instrument suite.

An important aspect of the Endurance concept study was that the study compared two distinct implementation options:

- **Endurance-A** (A is for Artemis astronaut): The Endurance-A rover would land in the center of SPA, and traverse ~2,000 km out of SPA through Poincaré and Schrödinger basins. The baseline[6] mission would collect a total of 100 kilograms of samples from 12 sites. (The threshold[7] mission would collect 1.2 kilograms from 6 sites.) Endurance-A would deliver those samples to Artemis astronauts who could return that large sample mass to Earth.

- **Endurance-R** (R is for Robotic): The Endurance-R rover would land in the Poincaré basin, and traverse ~1,750 km across the center of SPA, before heading northwest towards Apollo basin. The baseline mission would collect a total of 2.4 kilograms from 12 sites. (The threshold would collect 1.2 kilograms from 6 sites.) Endurance-R would deliver those samples to a robotic Earth Return Vehicle (ERV) which receives the samples and delivers them to Earth.

The only major differences between the Endurance-A and Endurance-R concepts were the traverse (see **Figures 2–3**), sample mass, and sample collection system. Endurance-A's traverse was highly constrained by the need to rendezvous with Artemis astronauts at the lunar south pole, whereas Endurance-R's traverse was free to prioritize sample sites purely by scientific priority. (Note: Endurance-A's traverse constraints could be substantially lessened if it could rendezvous with a non-polar Artemis mission in SPA—but this was deemed beyond the scope of the Decadal Survey study.) For sampling, Endurance-A used a simple scoop to collect large sample masses (100 kg total), whereas Endurance-R used a more complex sieving system to collect small samples and deposit them in a sample canister capable of fitting within an OSIRIS-REx–like sample return capsule (2.4 kg total sample mass). The Decadal Survey ultimately only recommended Endurance-A for implementation, owing to its lower cost and higher science return compared to the fully robotic Endurance-R approach. Since only Endurance-A was recommended for implementation, we simplify the name to just "Endurance."

Endurance's current sample requirements are described in **Table 3**. One of the consequences of Endurance's complex inception is that there is a substantial difference in sample requirements between the baseline mission and threshold mission. The threshold sample requirements (1.2 kg total returned sample mass) were designed to be consistent with the Endurance-R concept (where the samples must fit in a sample return capsule which has very limited mass capability), and meet the absolute minimum necessary sample mass requirements based on a grassroots estimate of the laboratory analyses required to accomplish Endurance's science objectives (see Appendix B of the Endurance concept study report[2]). The baseline sample requirements (100 kg total returned sample mass) were designed to be ambitious, based on analogy to Apollo sample return masses, and drive the rover design to be extremely capable. A future SDT (**Box 1**) is necessary to revisit these sample requirements and define more precise baseline and threshold requirements (see also **Finding #1**).

---

[6] A "baseline science mission" is the mission that, if fully implemented would fulfill the baseline science requirements which are necessary to achieve the full science objectives of the mission.

[7] A "threshold science mission" is a descoped baseline science mission that would fulfill the threshold science requirements which are necessary to achieve the minimum science acceptable for the investment.

| INSTRUMENT | KEY PARAMETERS |
|---|---|
| **TriCam: Stereo Imagers** | Pair of color stereo imaging cameras to characterize local geology.<br>− 3 color bands: Red, Green, Blue (RGB)<br>− Field of View (FOV): 50° × 37.5° (>180° with mosaicking)<br>− Instantaneous Field of View (IFOV): 220-µradians (1-mm pixel scale at 4-m range, 1-cm pixel scale at 45-m range) |
| **TriCam: FarCam** | Monochromatic narrow angle camera for long-range reconnaissance.<br>− FOV: 6.7° × 5° (>180° with mosaicking)<br>− IFOV: 50-µradians (5-cm pixel scale at 1-km range) |
| **Point Spectrometer (PS)** | Near-infrared spectra for determining mineralogy.<br>− 16 color bands (300-nm to 1,400-nm)<br>− FOV: <0.3° (3-m spot size at 100-m range) |
| **Hand Lens Imager (HLI)** | Hand lens imager for imaging lunar regolith and rocks at the microscopic scale.<br>− 3 color bands (RGB)<br>− 2-cm to infinite focal range (15-µm pixel scale at 23-mm range)<br>− Active focus and illumination (for day and night operations) |
| **Alpha Particle X-Ray Spectrometer (APXS)** | X-ray spectra for determining elemental abundances.<br>− Energy range: 0.4-MeV to 10-MeV<br>− Sensing depth: 2-cm (below space weathering rind) |
| **Magnetometer** | Magnetometer for measuring local magnetic field.<br>− Magnetic field intensity: ±100,000-nT<br>− Precision: 0.2-nT<br>− Sampling rate: 1-Hz |
| **Gamma Ray and Neutron Spectrometer (GRNS)** | Gamma-ray and neutron spectra for determining elemental abundances.<br>− Energy range: 0.4-MeV to 10-MeV<br>− Sensing depth: 30-cm |
| **Automated Radiation Measurements for Aerospace Safety (ARMAS)** | Radiation monitor for measuring heavy ions, alphas, protons, neutrons, electrons, and gamma-rays.<br>− Absorbed energy: 60-keV to >15-MeV |
| **Electrostatic Analyzer** | Radiation monitor to characterize solar wind ions and other ionizing radiation.<br>− Energy range: 200-eV to 20-keV<br>− Energy resolution: 8% |
| **Laser Retroreflector** | Passive laser retroreflector for geodetic measurements from another spacecraft. |
| **Inertial Measurement Unit (IMU)** | LN-200S accelerometer, for measuring gravitational accelerations.<br>− Sensitivity: 10 mGal |

*Table 2: Endurance's baseline instrument suite, as defined by the Endurance concept study report[2]. Note that one of the ground rules for the Endurance concept study was that it was to utilize the same instrument suite as the Intrepid instrument suite[3]. A future SDT (**Box 1**) is necessary to further define Endurance's instrument and measurement requirements.*

| **ENDURANCE SAMPLE REQUIREMENTS** | | |
|---|---|---|
| | **THRESHOLD:** | **BASELINE:** |
| **Sample size/mass per sample site:** | 200 grams of regolith, including ≥20 rocks between 0.5–2.0 centimeters in diameter ("rocklets"), per sample site | 8 kilograms of regolith and rocks per sample site |
| **Total number of sample sites:** | 6 sample sites | 12 sample sites |
| **Total sample mass:** | 1.2 kg (200 grams per site × 6 sites) | 100 kg (8 kilograms per site × 12 sites) |

*Table 3: Endurance's threshold and baseline sample requirements as defined by the Endurance concept study report[2]. A future SDT (**Box 1**) is likely necessary to further refine these requirements.*

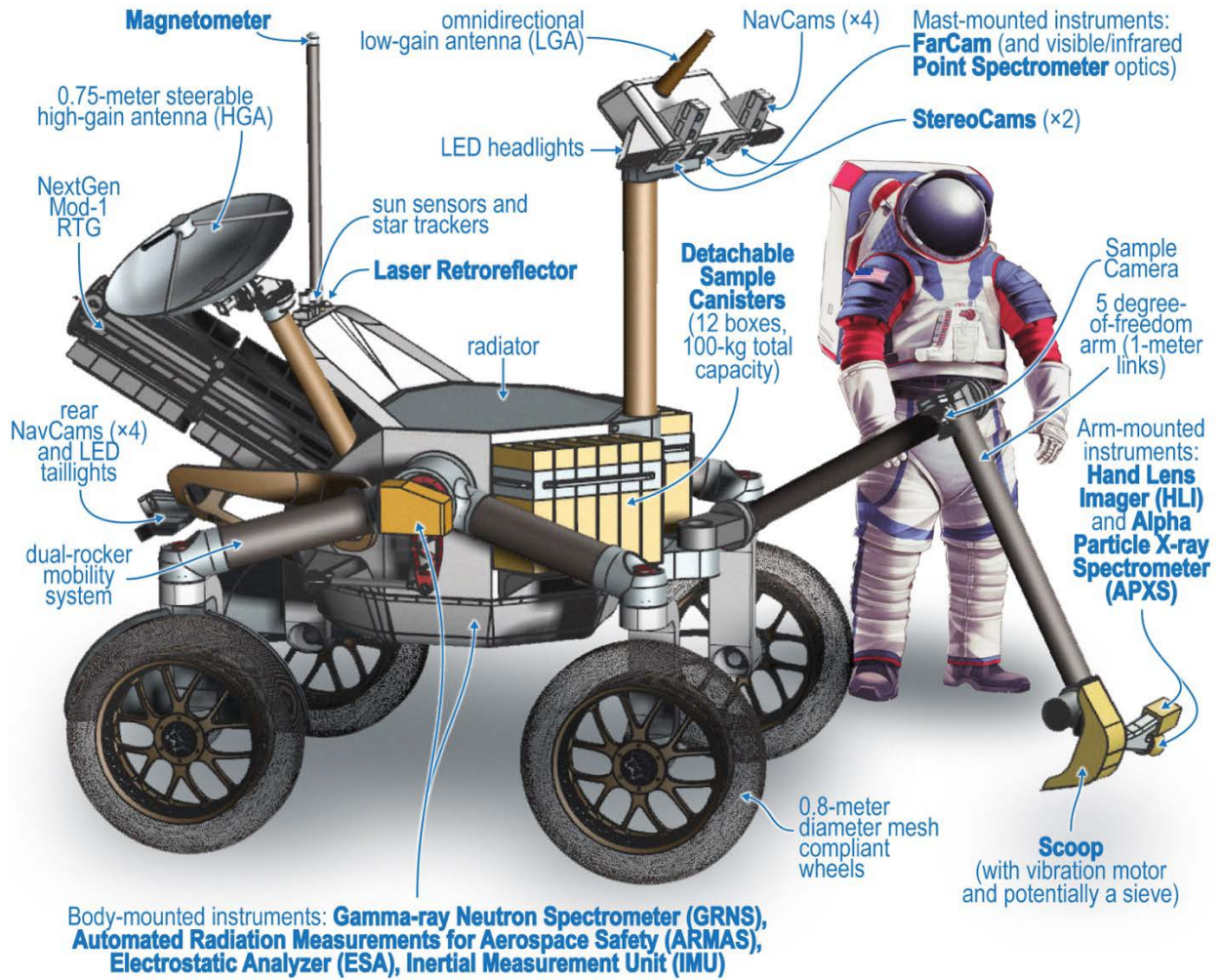

***Figure 4: Artist concept of the Endurance rover, as developed during the Decadal Survey concept study report[2].*** *Science instruments and hardware are identified in bold. A notional Artemis astronaut is shown for scale behind the rover.*

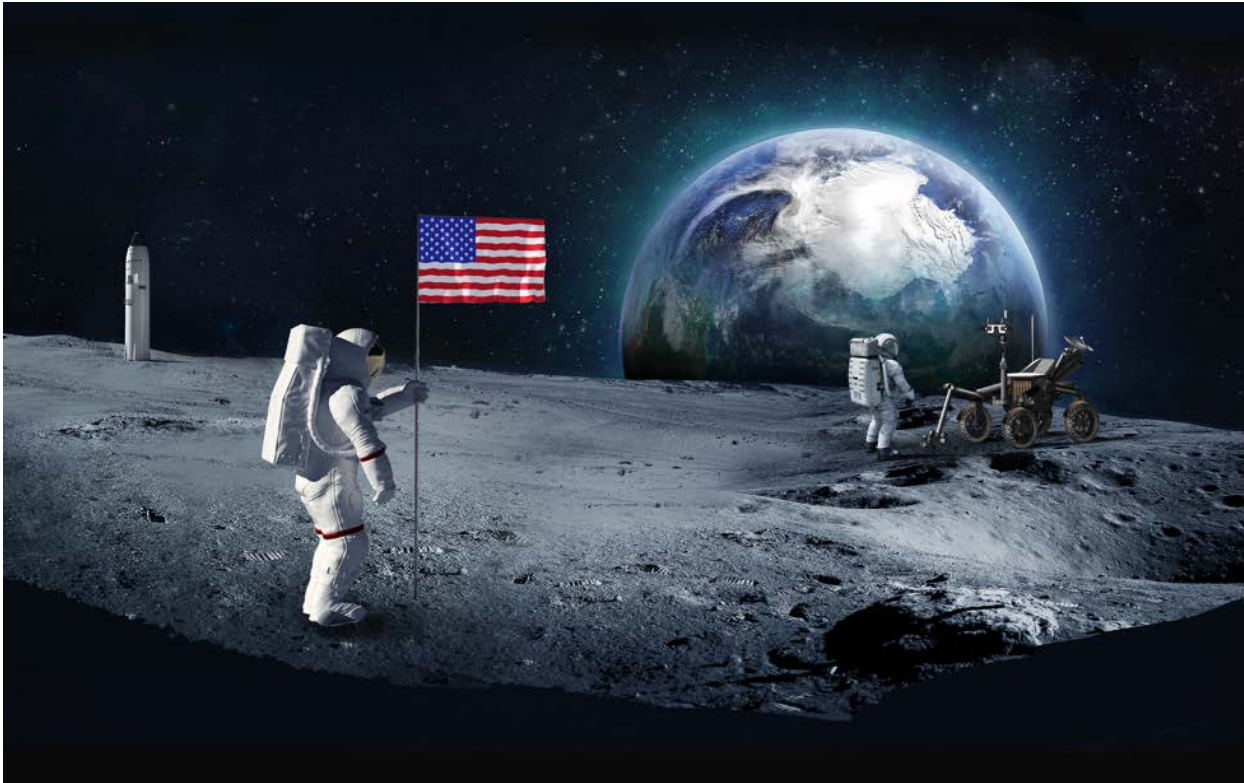

*Figure 5: Artist concept of Endurance rendezvous with Artemis astronauts.* Figure adapted from the Endurance concept study report[2]. Artwork by David Hinkle.

## 2.3  THE PROGRAMMATIC CONTEXT FOR ENDURANCE

### 2.3.1  ENDURANCE IN THE DECADAL SURVEY

The *Origins, Worlds, and Life* Decadal Survey[1] recommended Endurance for implementation as a medium-sized (New Frontiers-class) strategic mission, and the highest priority of NASA's Lunar Discovery and Exploration Program (LDEP):

> *"Recommendation: Endurance-A should be implemented as a strategic medium-class mission as the highest priority of the Lunar Discovery and Exploration Program. Endurance-A would utilize CLPS to deliver the rover to the Moon, a long-range traverse to collect a substantial mass of high-value samples, and astronauts to return them to Earth."* (Decadal Survey, 22-17)

The Decadal Survey emphasized the synergies between Endurance and Artemis, and benefits for the Planetary Science Directorate (PSD):

> *"Return of Endurance-A samples by Artemis astronauts is the ideal synergy between NASA's human and scientific exploration of the Moon, producing flagship-level science at a fraction of the cost to PSD through coordination with Artemis."* (Decadal Survey, 22-17)

The Decadal Survey's prioritization of Endurance was reflected in its final budget recommendations, including recommending Endurance for a mid-decade start in both the aspirational "Recommended" budget and more modest "Level" budget. In the event that the budget is reduced below the "Level" budget, the Decadal Survey provides prioritized budgetary decision rules for how to adapt. Endurance was far down the descope list (item 5), coming after delaying the start of the Uranus Orbiter and Probe mission, and reducing the number of Discovery and New Frontiers options in the coming decade.

### 2.3.2 ENDURANCE AFTER THE DECADAL SURVEY

After the release of the *Origins, Worlds, and Life* Decadal Survey, NASA provided an initial response within 90-days. With regards to the Decadal Survey's recommendation to implement Endurance, NASA stated:

*"NASA acknowledges this recommendation. NASA is assessing the feasibility of Endurance-A and considering different implementation approaches to achieving Endurance-A's science objectives within the scope of programmatic and budgetary guidance provided by this decadal survey. NASA is considering all solutions, including alternative solutions that emerge as the capabilities of the CLPS vendors and the Artemis campaign evolve."* (NASA's response to the Decadal Survey[8])

The Lunar Exploration Analysis Group (LEAG) responded to the Decadal Survey and NASA's response in a finding in the first LEAG meeting after the release of the Decadal Survey. LEAG reaffirmed the importance of SPA and the Endurance concept, and encouraged NASA to actively involve the science community in its development:

*"LEAG renews its emphasis on the scientific importance of the collection and return to Earth of samples from the South Pole–Aitken basin (SPA) to address a long-standing top priority for lunar and planetary science, and we encourage NASA to continue to explore and openly communicate options for implementation. Despite the continued recognition of the importance of this sample collection and analysis, a mission to accomplish this has not yet been selected through New Frontiers. The recent Origins, Worlds, and Life Decadal Survey identified a new mechanism to accomplishing SPA sample return through the Endurance mission concept, which was the highest rated mission in the Survey. We strongly encourage NASA to actively involve the planetary science community as it considers how best to implement the Endurance concept. In addition, we urge regular updates to the community regarding study results of budget, implementation options, and timeline. We look forward to seeing how the out-of-the-box thinking of the Endurance mission will finally begin to address this priority and we eagerly await news of the mission development."* (LEAG finding[9])

In early 2023, NASA tasked the Jet Propulsion Laboratory (JPL) with performing a follow-on study for Endurance, including examining trades that may reduce the total cost of the mission. This work is ongoing, but multiple trades are being considered including: (1) simplifying the rover mobility system; (2) reducing the number of instruments to only those required for sample collection and geologic context; (3) evaluating an option using solar power; and (4) investigating the feasibility of serviceability by astronauts, including accommodation of modular instruments, and replacing key hardware (*e.g.*, wheels). JPL is also assessing alternative implementation approaches (*i.e.*, concepts that do not necessarily require a single long-range rover) to ensure that the Endurance science goals are achieved using the optimal architecture. The current study will continue into 2025.

---

[8] NASA's initial (90-day) written response to the 2023–2032 Planetary Science and Astrobiology Decadal Survey (2022): https://science.nasa.gov/science-pink/s3fs-public/atoms/files/Initial%20(90-day)%20written%20response%20to%20the%202023%e2%80%932032%20Planetary%20Science%20and%20Astrobiology%20Decadal%20Survey.pdf.

[9] 2022 Annual Meeting of the Lunar Exploration Analysis Group (LEAG): Summary and Findings: https://www.hou.usra.edu/meetings/leag2022/LEAG2022AnnualMeetingFindings_FINAL.pdf.

NASA's Exploration Science Strategy and Integration Office (ESSIO) is currently planning the initiation of a Science Definition Team (SDT) for Endurance, which would define the specific scientific requirements for the mission (**Box 1**). It is expected that the SDT will begin in late 2024 or early 2025.

## Box 1: The Role of a Science Definition Team (SDT)

*__What are SDTs?__ Science Definition Teams (SDTs) are the process by which NASA develops the science justification, science objectives, measurement requirements, and mission requirements for strategic missions. The most relevant, recent example of an SDT is the Mars 2020 SDT[10].*

*__Why do we need SDTs?__ Endurance, like many strategic missions, originated from a mission concept developed during a Decadal Survey. While these mission concept studies are thorough, they are primarily feasibility studies—proofs of concept—and do not necessarily provide the level of detail required to confidently begin to design a real mission. An SDT can develop the science traceability for a mission, including defining goals, objectives, measurement and mission requirements, etc. Additionally, as missions develop programmatic constraints may change, including cost and schedule, which affect mission development. An SDT can prioritize scientific investigations to allow NASA to maximize the science return within available resources. SDTs are also a critical opportunity for the scientific community to become involved in the mission concept, shaping the science and building community support for the concept.*

*__How do SDTs operate?__ SDTs are traditionally initiated by NASA and consist of 10~20 scientists and subject matter experts who, over the course of 6~12 months, develop a report that provides NASA with the science rationale, objectives, requirements, and traceability for the mission in question. SDTs usually work in concert with an engineering support team at a NASA lab (e.g., Jet Propulsion Laboratory, Goddard Space Flight Center, Applied Physics Laboratory) to develop an engineering point design to ensure that the science requirements are executable given available constraints (e.g., cost, mass, power, schedule). Upon completion of the SDT report, the teams are disbanded. SDTs are usually early in the mission formulation process (Pre-Phase-A or Phase-A), and may precede open competitions for instruments or science team membership.*

*__What is next for the Endurance SDT?__ NASA's Exploration Science Strategy and Integration Office (ESSIO) within the Science Mission Directorate (SMD) is planning to initiate a South Pole–Aitken basin sample Return and eXploration (SPARX) SDT in late 2024 or early 2025. It is expected that this report and the findings therein may be valuable input to the Endurance SDT.*

---

[10] Mars 2020 Science Definition Team report: https://mars.nasa.gov/mars2020/files/mars2020/SDT-Report%20Finalv6.pdf.

# 3. THE ENDURANCE SCIENCE WORKSHOP

## 3.1 GOAL OF THE WORKSHOP

The goal of the Endurance Science Workshop was to bring together the planetary science community—including scientists, engineers, technologists, commercial partners, and more—to discuss the science that could be accomplished by the Endurance mission concept.

## 3.2 WORKSHOP STRUCTURE AND STATISTICS

The Endurance Science Workshop was held at the Cahill Center on the California Institute of Technology campus, over three days (9–11 August 2023). **Figure 6** shows some photos from the workshop. The workshop was structured around a series of topical sessions, combining invited and solicited talks, and culminating in open panel discussion sessions. **Table 4** shows a summary of the program. The workshop was hybrid, with about 2/3rd of participants participating remotely (**Table 5**). Details about the workshop, including abstracts, can be found on the Endurance Science Workshop website[11].

| DAY 1<br>WEDNESDAY,<br>AUGUST 9TH, 2023 | DAY 2<br>THURSDAY,<br>AUGUST 10TH, 2023 | DAY 3<br>FRIDAY,<br>AUGUST 11TH, 2023 |
|---|---|---|
| Introduction to Endurance | Lessons Learned | Traverse |
| South Pole–Aitken Basin Overview | Science Measurements, Part 1 (Geophysics) | Rover Design |
| Sample Science | Science Measurements, Part 2 (Mineralogy, Chemistry, and Geology) | Closing Discussion |

*Table 4: Endurance Science Workshop program summary.*

---

[11] Endurance Science Workshop 2023 website: https://www.hou.usra.edu/meetings/endurance2023/.

| WORKSHOP STATISTICS | |
|---|---|
| Total number of abstracts submitted: | 66 |
| Total number of registrations: | 352 |
| Peak number of in-person attendees: | 55 |
| Peak number of virtual attendees: | 117 |
| Total number of speakers: | 64 |
| Total number of panelists: | 27 |
| Total number of rovers: | 1 (see **Figure 6**) |

*Table 5: Endurance Science Workshop by the numbers.*

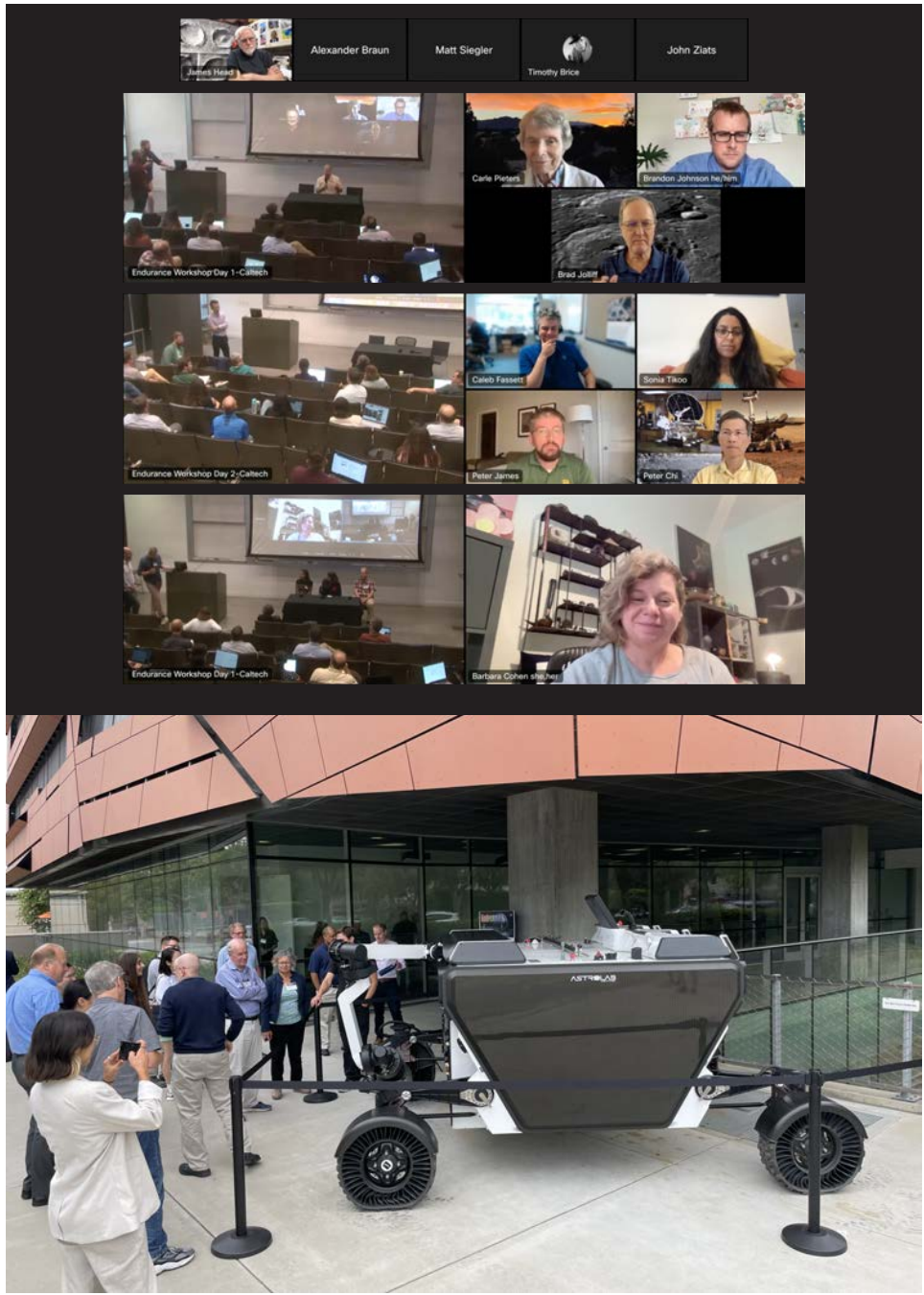

***Figure 6: Candid photos from the Endurance Science Workshop.*** *Top: screenshots from the Endurance Science Workshop livestream from panel discussions on Days 1 and 2. Bottom: photograph of the Astrolab FLEX rover testbed outside of the Cahill Center at the California Institute of Technology on Day 3 of the Endurance Science Workshop.*

# 4. FINDINGS OF THE ENDURANCE SCIENCE WORKSHOP

In this section, we summarize the major findings identified by the Endurance Science Workshop science organizing committee. Findings are listed in **Table 6**. Findings are not prioritized. In addition to this report, findings have been publicly presented at several conferences and meetings, including LEAG[12], AGU[13], LPSC[14], and CAPS[15].

| | ENDURANCE SCIENCE WORKSHOP 2023 FINDINGS | |
|---|---|---|
| 1 | Endurance is an exciting concept that would address long-standing, high-priority lunar and planetary science questions—and the community is ready for it. | Page 23–26 |
| 2 | Endurance's sample science objectives are achievable, although they would require coordinated analysis techniques and numerous diverse samples. | Page 27–30 |
| 3 | Geologic context is essential for addressing Endurance's science objectives. | Page 31–33 |
| 4 | While Endurance's objectives center on sample return, Endurance's long traverse would enable a variety of additional transformative science investigations. | Page 34–41 |
| 5 | Endurance is an ambitious mission that would be enabled and enhanced by investing in developing key technologies now. | Page 42–50 |
| 6 | Endurance should strive to include more diverse perspectives in its formulation—particularly from early-career scientists and engineers who will ultimately operate the rover and analyze the samples. | Page 51–52 |

**Table 6: Findings of the Endurance Science Workshop 2023.**

---

[12] Keane et al. (2023), Annual Meeting of the Lunar Exploration Analysis Group (LEAG): https://www.hou.usra.edu/meetings/leag2023/presentations/Wednesday/1630_Keane.pdf.

[13] Tikoo et al. (2023), Fall Meeting of the American Geophysical Union (AGU): https://ui.adsabs.harvard.edu/abs/2023AGUFM.P31B..08T.

[14] Keane et al. (2024), Lunar and Planetary Science Conference (LPSC): https://ui.adsabs.harvard.edu/abs/2024LPICo3040.2172K.

[15] Keane et al. (2023), presentation to the Committee on Astrobiology and Planetary Sciences (CAPS) autumn meeting: https://www.nationalacademies.org/event/10-24-2023/committee-on-astrobiology-and-planetary-sciences-autumn-2023-meeting.

| FINDING #1 | Endurance is an exciting concept that would address long-standing, high-priority lunar and planetary science questions—and the community is ready for it. |
|---|---|

Sample return from SPA has been the top priority in the last three Planetary Science (and Astrobiology) Decadal Surveys[16,17,1], and a multitude of other strategic documents produced by the lunar and planetary science community[18,19,20,21]. SPA sample return is highly prioritized because of the broad scientific ramifications expected from analyzing samples from the deepest and oldest basin on the Moon. As the oldest impact basin on the Moon, SPA provides an opportunity to understand the frequency of giant impacts early in the Solar System's history (when life was first emerging on Earth), including testing the long-debated "Late Heavy Bombardment" hypothesis. As the largest and deepest impact basin on the Moon, SPA provides an opportunity to study materials exhumed from the Moon's lower crust and mantle that are largely obscured or inaccessible elsewhere. Understanding the interior structure and composition of the Moon and how it changed with time is important to understanding how other rocky worlds evolved. The science motivation for SPA exploration is further detailed in **Section 2.1**.

The high priority of SPA is reflected in the substantial amount of research that has gone into studying SPA. Per NASA's Astrophysical Data System[22], there are 1,000+ refereed publications about SPA in the literature, and an additional 1,100+ conference publications about the topic over the last several decades (**Figure 7**). Recent work includes new analyses enabled by a variety of new robotic missions (both orbital and landed), new theoretical and modeling developments, and new revelations from continued analysis of lunar samples using state-of-the-art laboratory techniques. Furthermore, there have been numerous efforts to propose missions to accomplish the goals of SPA sample return, such as the MoonRise mission concept[23–24] which was a finalist for the NASA's third New Frontiers opportunity. These concept studies and missions have advanced both the scientific and technical understanding necessary for accomplishing the goals of SPA sample return.

---

[16] National Academies of Science: New Frontiers in the Solar System: An Integrated Exploration Strategy (2003): https://nap.nationalacademies.org/catalog/10432/new-frontiers-in-the-solar-system-an-integrated-exploration-strategy.

[17] National Research Council of the National Academies: Vision and Voyages for Planetary Science in the Decade 2013-2022 (2011): https://nap.nationalacademies.org/catalog/13117/vision-and-voyages-for-planetary-science-in-the-decade-2013-2022.

[18] Lunar Exploration Analysis Group (LEAG): Lunar Exploration Roadmap: https://www.lpi.usra.edu/leag/roadmap/.

[19] National Research Council of the National Academies: The Scientific Context for Exploration of the Moon (2007): https://nap.nationalacademies.org/catalog/11954/the-scientific-context-for-exploration-of-the-moon.

[20] Lunar Exploration Analysis Group (LEAG): Advancing Science of the Moon: Report of the Lunar Exploration Analysis Group Special Action Team (2018): https://www.lpi.usra.edu/leag/reports/ASM-SAT-Report-final.pdf.

[21] NASA: Artemis III Science Definition Team Report (2020): https://www.nasa.gov/wp-content/uploads/2015/01/artemis-iii-science-definition-report-12042020c.pdf.

[22] Astrophysical Data System (ADS): https://ui.adsabs.harvard.edu/.

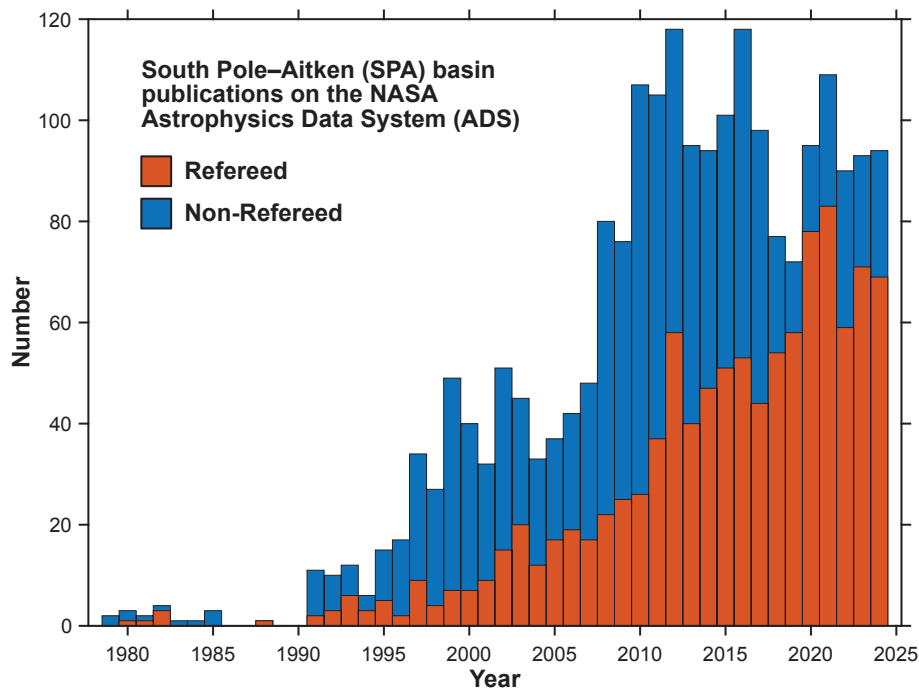

*Figure 7: There have been over 2,000 publications about South Pole-Aitken.* This data is provided by the NASA Astrophysical Data System (ADS), and shows all publications that include "South Pole–Aitken" either in the title, abstract, or main body of the manuscript.

The Endurance Science Workshop both reaffirmed the scientific merit of sample return from SPA, and demonstrated that the lunar and planetary science community has already done a lot of the necessary work to enable the Endurance mission concept. Put simply: the community is ready. While there were debates about specific science implementation trades (*e.g.*, sample size requirements, traverse options, instrument requirements) that will require further study (*e.g.*, via an SDT), there was no substantial

---

[23] Jolliff et al. (2017), Lunar and Planetary Science Conference: https://www.hou.usra.edu/meetings/lpsc2017/pdf/1300.pdf.

[24] Jolliff et al. (2021), white paper for the 2023–2032 Planetary Science and Astrobiology Decadal Survey: https://doi.org/10.3847/25c2cfeb.5309cd69.

debate about the science motivating the Endurance concept.

Since the Endurance Science Workshop, China successfully returned the first samples from the Apollo basin in northern SPA with the Chang'e 6 mission. Chang'e 6 landed on mare deposits within the Apollo basin, and returned ~1.9 kilograms of regolith samples. Analysis of the Chang'e 6 samples is ongoing, but it is expected that Chang'e 6 may contribute to addressing some questions about the nature of lunar farside volcanism[25,26,27] relevant to Endurance objective 4 (**Table 1**). Preliminary analyses suggest that the Chang'e 6 samples are similar to mare basalts collected by Apollo[28,29]. However, Chang'e 6's landing site is far from the central portions of SPA where the crust is thinnest and there is more geologic evidence for SPA impact melt and deeply-excavated material—all of which are likely necessary to address Endurance's broad science objectives (**Table 1**). This is shown graphically in **Figure 8**. Nonetheless, Chang'e 6 can provide important information that can feed forward into the development of Endurance. Similarly, future samples returned from the south pole (on the southern rim of SPA) by Artemis or future Chang'e missions may provide additional insights. However, sample return missions from single locations will only ever provide individual "pieces of the puzzle" and have challenges addressing Endurance's broad science goals. Endurance is really a sample return campaign, coupled with the in-situ investigations afforded by a large rover and long traverse, which is hard to fully capture with individual sample return missions from single sites.

---

[25] Zeng et al. (2023), Nature Astronomy: https://doi.org/10.1038/s41550-023-02038-1.

[26] Qian et al. (2024), Astrophysical Journal Letters: https://doi.org/10.3847/2041-8213/ad698f.

[27] Jia et al. (2024), Icarus: https://doi.org/10.1016/j.icarus.2024.116107.

[28] Li et al. (2024), National Science Review: https://doi.org/10.1093/nsr/nwae328.

[29] Preliminary analysis by Randy Korotev: https://sites.wustl.edu/meteoritesite/items/the-chemical-composition-of-lunar-soil/.

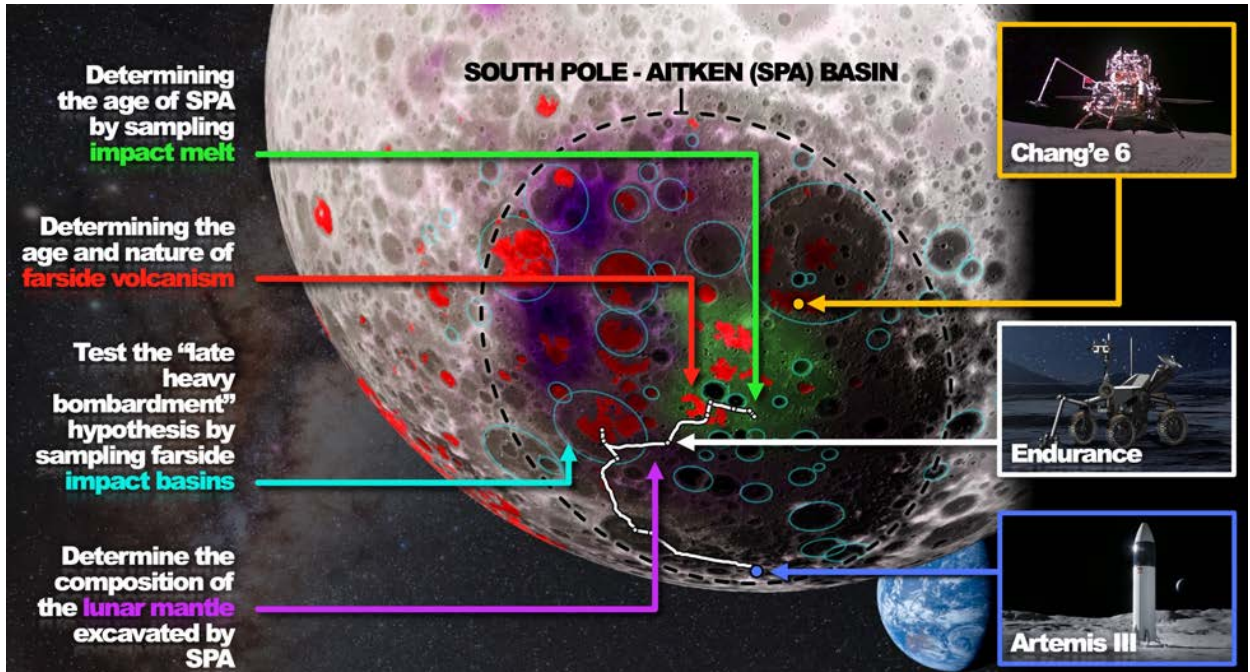

*Figure 8: Endurance would traverse and sample diverse terrains across South Pole–Aitken (SPA), beyond what is possible from sample return missions to individual sites.* Topographic map of the farside of the Moon, centered on the SPA basin, enclosed in the dashed ellipse. Topography ranges from −7 km (black) to +7 km (white). SPA is a geologically diverse terrain on the Moon, and it is necessary to sample multiple terrains to address Endurance's high-priority, long-standing science objectives. Different geologic terrains are indicated by different colors: green indicates the SPA Compositional Anomaly (SPACA), which may be related to SPA impact melt; red indicates farside mare basalts; cyan ellipses enclose pre-Imbrian impact basins; purple indicates regions with enhanced thorium abundance, which may relate to mantle material. Chang'e 6 (yellow point) and Artemis III (blue point) have/will acquire samples from localities on the margins of SPA, whereas Endurance's 2,000 km traverse (white line) would acquire samples and data from within the core of SPA—where the highest priority samples are accessible. Figure adapted from the Endurance concept study report[2].

| | |
|---|---|
| **FINDING #2** | **Endurance's sample science objectives are achievable, although they would require coordinated analysis techniques and numerous diverse samples.** |

Endurance's science objectives center on the analysis of returned samples. While there have been substantial advances in sample analysis capabilities in recent years—from the analysis of new asteroid samples by missions like Hayabusa2 and OSIRIS-REx, new analytical techniques, and even the analysis of "new" preserved lunar samples from the Apollo Next Generation Sample Analysis Program (ANGSA)[30]—there was consensus at the Endurance Science Workshop that Endurance's sample science objectives will be challenging, but achievable. The Endurance Science Workshop identified several important topics with regard to the future selection, collection, and analyses of Endurance's returned samples which are detailed here. One of the most important topics for the forthcoming Endurance SDT (**Box 1**) will be to address these issues and carefully define Endurance's sample requirements. While the Endurance concept study report[2] defined sample science requirements, these requirements should be considered a starting point, and require more careful scrutiny by the scientific community. **Endurance will require a large number and significant mass of diverse samples from across the SPA basin and its surrounding terrains in order to confidently address the science objectives defined by the Decadal Survey.**

Endurance's science objectives involving geochronology—particularly Objective 1 (determine the age of SPA) and Objective 2 (determine the age of other large farside impact basins) (**Table 1**)—may be the most challenging to address. These basins are ancient, with approximately four billion years of subsequent impact bombardment and geologic processes that can obscure or obfuscate the true age of the basins. Nevertheless, multiple approaches exist to mitigate this confusion. First, it is desirable to collect samples from multiple geographically and geologically distinct locations across SPA and its surrounding terrains. Diverse samples are advantageous because the analysis of an ensemble of diverse samples can help build a more holistic geologic picture. This is a major strength of Endurance (which nominally returns samples from 12 sites across SPA) compared to other sample return missions and concepts that return sample from only one location. Second, collecting a large number of samples of significant mass means that some sources of confusion can be mitigated via statistics. Rather than depend on analyses of a single rock or regolith sample, scientific hypotheses can be tested against ensembles of samples using a variety of methods.

After billions of years of impacts, the regolith in SPA will contain rocks altered and transported by multiple impact craters, potentially complicating the identification of their source impact crater. The more rocks collected from different geographic locations makes it more likely that Endurance will sample distinct impact events and unravel this geologic story. A large cache of samples also increases the probability of collecting unexpected or surprising samples that may be particularly informative. An implicit implication of this discussion is that **the threshold sample mass requirement identified for Endurance in the concept study report (1.2 kilograms, Table 3) may be inadequate to meet the science requirements. The true threshold sample mass may be larger (10s of kilograms), although this requires verification by the SDT (Box 1).**

---

[30] Shearer et al. (2024), Space Science Reviews: https://doi.org/10.1007/s11214-024-01094-x.

Collecting a large amount of samples is advantageous for multiple additional reasons. For comparison, imagine a traditional sample return mission from a single site. In this hypothetical case, only a small amount of material (~1 kilogram) may be collected. The regolith in a single sample location may or may not contain the ideal material sufficient for testing only one or a few motivating science hypotheses. NASA curation would also likely withhold >50% of the sample according to curation and analysis protocols. By returning a large sample cache, Endurance ensures many more scientific investigations, by many more teams, substantially enhancing the scientific impact of the mission.

Our experience with analyses of lunar samples over the past 50+ years has taught us that determining the age of an impact event on the Moon can be challenging. Multiple geochronometers (*e.g.*, Rb-Sr, U-Pb, K-Ar) would ideally be applied to a given impactite sample to provide the correct interpretation for when the impact occurred, and how the rock was affected by subsequent impacts and geologic events. Impactites are rocks created, or modified, by the intense heat and shocks associated with impacts. Different geochronometers can yield different ages as they respond to thermal disturbances (*e.g.*, impacts) differently depending on isotopic system closure temperatures and each element's diffusivity in the sample[31,32,33]. Thus, it is also necessary to understand the petrology (*i.e.*, mineralogy, texture) of the sample in order to understand the rock's formation and evolution and to properly interpret ages produced by different geochronometers. Different isotopic systems also require different masses of material[34]. Endurance's sample science objectives will require coming at those objectives from multiple directions, using multiple analytical techniques, on multiple samples. Interpretation of age information also highlights the importance of geologic context (discussed further in **Finding #3**).

For determining the ages of impact basins, some of the best samples are impact melt rocks or impact melt breccias. Impact melt is a once-molten rock that formed from the that formed from the intense heat and pressure of an impact. A breccia is a conglomerate consisting of rock and mineral fragments in an (sometimes devitrified) impact-melt matrix or in a fine-grained crystalline or fragmental matrix. In an ideal case, an impact creates melt that can crystallize into a newly formed rock and several geochronometers (*e.g.*, Ar-Ar, Rb-Sr, U-Pb) are reset and begin their decay (*i.e.*, they "start ticking"). However, these types of samples are less common than impact melt breccias, which are often complex—containing components from many sources, delivered by multiple impact events. Therefore, larger mass samples (*e.g.*, >25 grams, roughly equivalent to a 2.5-centimeter diameter rock) is desired to maximize the probability of identifying suitable rocks and lithologies, and allow the use of multiple chronometers on multiple components within the rock (matrix, clasts, etc.) to decipher its assembly history. Rocks of this size are at the upper-edge of Endurance's current baseline sample requirements (0.5–2.0 centimeters, **Table 3**). **The future SDT (Box 1) should carefully consider the sample size requirements, and in particular, refine the required size range for rocks, and the number of rocks to be collected from each sample site.**

---

[31] Borg et al. (2015), Meteoritics & Planetary Science: https://doi.org/10.1111/maps.12373.

[32] Gaffney et al. (2011), Meteoritics & Planetary Science: https://doi.org/10.1111/j.1945-5100.2010.01137.x.

[33] Borg & Carlson (2022), Annual Review of Earth & Planetary Sciences: https://doi.org/10.1146/annurev-earth-031621-060538.

[34] Shearer & Borg (2006), Geochemistry: https://doi.org/10.1016/j.chemer.2006.03.002.

Not all sample sizes have the same scientific value. Larger rocks (2-centimeters and larger, roughly the size of a US quarter dollar coin) are particularly important for many of the geochronology objectives and, in the case of impact breccias, for showing relationships between lithologies at a given site and between different sites. Larger samples are also advantageous as they enable additional laboratory analysis techniques, including preparation of multiple thin sections necessary for mineralogical and petrological characterization analyses, and potential specialized analyses such as impact shock barometry, rock magnetism, and characterization of physical properties. Larger rocks may be desirable, but there is a complex scientific and technical trade involved with defining the sample requirements. For example, if Endurance's sample return volume is fixed, increasing the size of required rocks would also imply a reduction in the total number of rocks, and potentially a more complex sample acquisition scheme. Taken to an extreme, if the required rock size increases dramatically, there is likely a point where the entire sample acquisition and storage strategy would need to be redesigned—going from a simple scooping and sieving approach to something more complex (*e.g.*, a "claw" to pick up individual rocks). One of the central challenges is that rocks on the Moon follow an exponential size-frequency distribution, with exponentially fewer larger rocks. Thus, if the required rock size increases, more regolith will have to be scooped and sieved to collect the same number of rocks. In the workshop and subsequent work[35], Paul Warren presented new analyses of the size-frequency distribution of Apollo regolith data. Based on the antiquity of SPA, he proposed that the regolith from the Apollo 16 mission to the ancient lunar highlands would be the best analog for the material that Endurance may encounter. Based on his work, he showed that only a small fraction (<1 weight percent) of regolith is made up of rock between 2–5 centimeters. This implies that **Endurance must sieve samples in order to maximize the number of rock in the sample cache. Scooping unsieved regolith alone is insufficient to meet the sample requirements without processing excessive volumes of lunar regolith.**

While larger rocks are a high priority, it is important not to neglect the value of having bulk regolith—including regolith fines (particles <0.5-centimeters in diameter, roughly equivalent to terrestrial sand or silt). Substantial advances in analysis techniques make even very small or fine-grained materials important for analyses, such as the collection and dating of zircons in lunar regolith[36], and volcanic and impact glass beads like those returned from Apollo, or most recently by Chang'e 5[37]. While rocks may be the priority, the workshop participants concurred that **Endurance should collect at least some unsieved regolith.** There was a notion at the workshop that Endurance should collect the high priority rocks first, and then fill any remaining sample container volume with unsieved regolith.

---

[35] Warren et al. (2024), Lunar and Planetary Science Conference: https://www.hou.usra.edu/meetings/lpsc2024/pdf/2512.pdf.

Finally, there was notable discussion at the workshop about the balance between collecting targeted samples that meet certain criteria, versus collecting samples representative of typical regolith. Stated another way, should Endurance seek out the "exceptional versus the exemplar"? There may be value in collecting random samples along the traverse, in addition to the more targeted collections from special sites.

Taken together, discussion at the Endurance Science Workshop highlighted the importance for the SDT (**Box 1**) to revisit Endurance sample requirements and sample acquisition strategy. This will likely entail a rich trade space analysis for optimizing the science value as a function of the number, mass, and type of samples, the number of sample sites, the use of in situ instruments, astronaut triage, etc.

---

[36] Borg et al. (2014), Meteoritics & Planetary Science: https://doi.org/10.1111/maps.12373.

[37] Wang et al. (2024), Science: https://doi.org/10.1126/science.adk6635.

| FINDING #3 | Geologic context is essential for addressing Endurance's science objectives |

Anytime a geologist is handed a rock and asked to identify it, one of the first questions the geologist might ask is "where did it come from?" Geologic context—or the observations and understanding of the geology and environment from which samples are acquired—is crucial for properly identifying, characterizing, and studying returned samples. Without geologic context, our ability to understand the scientific importance of a sample is severely hindered. This is the plight of lunar meteorites; while we know they originate from the Moon, we lack any geologic context, making it generally impossible to tie those meteorites to specific geologic features or processes, and limiting our ability to utilize these samples to their fullest scientific extent.

Geologic context is something that can be provided using a variety of datasets before, during, and after a mission. Consider, for example, the Apollo missions. Geologic context for the Apollo returned samples was accomplished using a variety of methods before, during, and after each mission. Before each mission, geologic context was provided by analysis of telescopic and spacecraft datasets of candidate landing sites. During each mission, astronauts provided geologic context by documenting their environment verbally, with photographic images, and other in situ measurements. After the missions, the geologic context for the Apollo samples has been continually refined by reanalysis of the Apollo data, and subsequent observations from new spacecraft (Lunar Prospector, Lunar Reconnaissance Orbiter, Chandrayaan-1, etc.). This is no different from Endurance.

**Geologic context is clearly important for addressing Endurance's science objectives.** Some of the geologic context for Endurance is already established by existing datasets (*e.g.*, Lunar Prospector, Lunar Reconnaissance Orbiter, Chandrayaan-1), and may be further refined by forthcoming missions (*e.g.*, Lunar Trailblazer). However, a key question remains: What is the *essential* geologic context that Endurance must acquire with its in-situ instruments while on the surface of the Moon? Simply landing on the Moon and grabbing samples agnostic to their geologic context is not sufficient for Endurance. However, Endurance will have limited resources—including cost, mass, power, data— which will translate to limited ability for in situ instruments and measurements. This makes it critical to define true threshold science requirements for in situ measurements. It is expected that the forthcoming SDT (**Box 1**) will define these requirements. Nonetheless, there was substantial discussion about the importance of various aspects of geologic context at the Endurance Science Workshop.

Geologic context for Endurance can span a range of physical scales, as illustrated in **Figure 9**. At the largest scales, we have the geology of the Moon and the impact basins that Endurance would traverse. At these large scales, it becomes important to contextualize Endurance's observations within what is observable from orbit. Field studies such as Endurance provide a fundamentally different perspective from orbital remote sensing datasets. There is an intrinsic synergy between these perspectives that allow both to be more than the sum of the components that is best leveraged when the surface and orbital observations are intentionally complementary. In some cases, this can be seen as simply making similar observations at higher spatial resolutions and different observation geometry. For example, visible/infrared spectra from the surface could provide information about chemistry, mineralogy, and geology down to centimeter-scale—commensurate with the scale of Endurance's scoop or workspace—whereas orbital data (*e.g.*, Moon Mineralogy Mapper ($M^3$) on Chandrayaan-1 and High-resolution Volatiles and Minerals Moon Mapper ($HVM^3$) on the forthcoming Lunar Trailblazer mission) can only map the lunar surface at 10s to 100s of meters per pixel which is sufficient to plan Endurance's traverse, but insufficient to fully contextualize centimeter-scale samples. These measurements can also be critical for verifying that the rover is in the geologic unit that you expected based on orbital data, and accurately mapping geologic contacts at the scale of the rover and sampling methodology. In other cases, some instruments are uniquely suited for collection of data on the surface rather than from orbit. For example, in-situ geochemical investigations like Alpha Particle X-Ray Spectroscopy (APXS) or Laser Induced Breakdown Spectroscopy (LIBS) only work in close proximity to the surface. A combination of approaches is essential for selecting locations with the highest probabilities of producing samples required to address the science objectives.

Geologic context also extends down to the smallest scales—even within individual samples. Many of the laboratory methods necessary for accomplishing Endurance's science objectives (*e.g.*, geochronology) require detailed understanding of the geologic context of those analyses at the scale of individual rocks, grains, and even crystals therein (see also **Finding #2**). Rock lithology, mineralogy, chemistry, and texture are all important observations that can inform the interpretation of laboratory measurements. In the baseline Endurance concept study[2], Endurance would utilize its in situ remote sensing instruments, including an APXS and Hand Lens Imager (HLI), to provide some micro-scale geologic context for samples. While there was consensus about the importance of mineralogical and chemical measurements, there was particular emphasis at the Endurance Science Workshop about the importance of sample texture. Mineralogy and chemical information alone are insufficient to determine whether a rock is a breccia or a basalt, or from the crust or mantle, which is critical for Endurance's science objectives related to the interior of the Moon (Objective 3 in **Table 1**). For volcanic rocks, texture can provide insight about the flow distance from the source, lava viscosity, lava temperature and cooling history, and emplacement conditions. While some of this analysis happens after the samples are returned home (*e.g.*, with photomicroscopy of samples), there was debate about how much of this could or should be performed while Endurance is on the surface of the Moon.

**There was substantial discussion at the Endurance Science Workshop about instruments and measurements that could provide geologic context. This discussion is continued in Finding #4. One of the major goals for the forthcoming Endurance SDT may be to define what constitutes essential geologic context, and what the associated measurement requirements are.** Defining these measurement requirements soon is important both to advance the rover design, and to prepare the

community for a possible future announcement of opportunity for competitively selected in-situ instruments for Endurance.

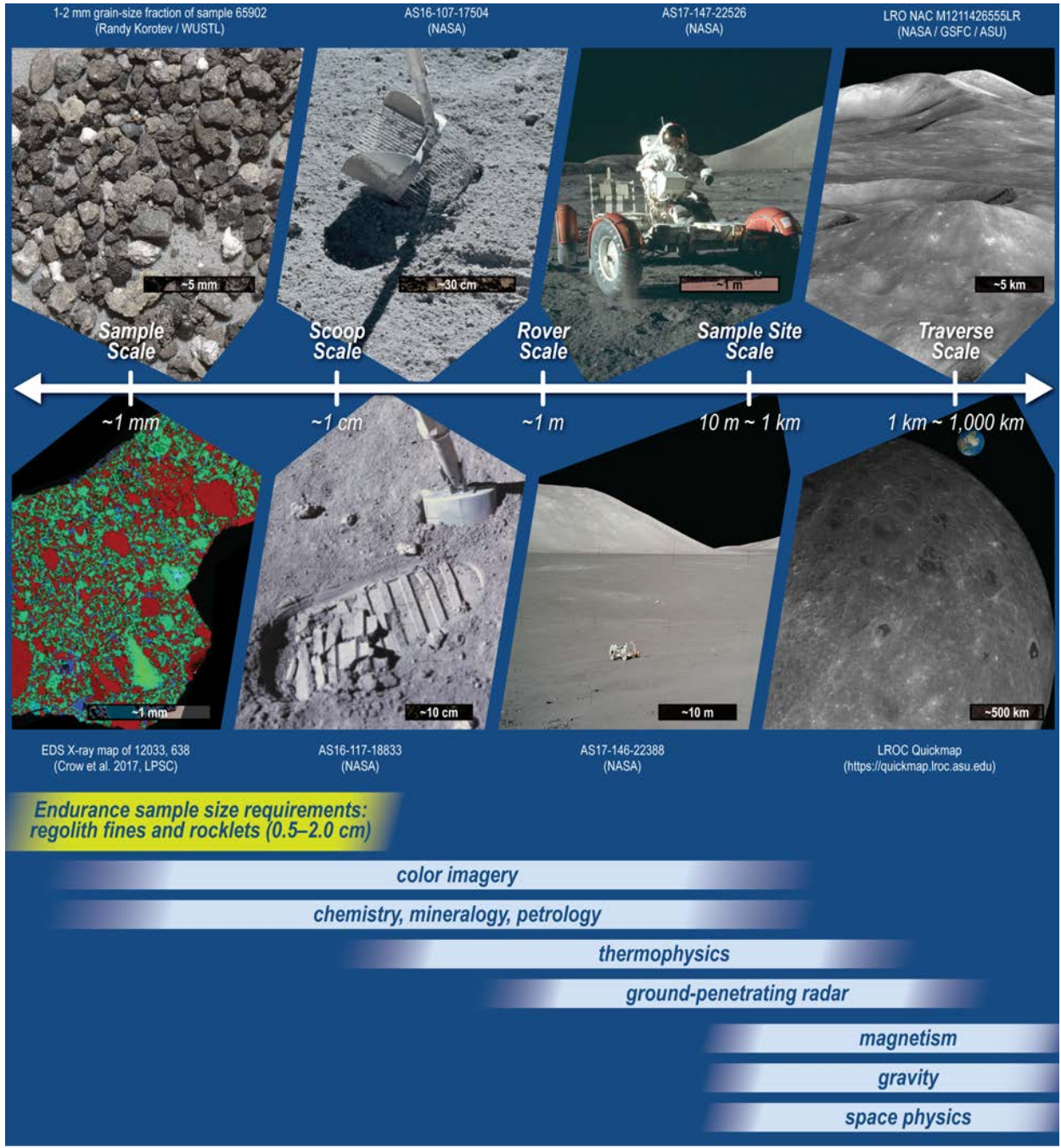

*Figure 9: Endurance will explore the Moon at a range of scales, from micro analyses of returned samples to geologic context at the scale of the rover and the traverse.* Different notional in situ measurements are shown across a range of relevant scales. Figure created by James Tuttle Keane.

| FINDING #4 | While Endurance's objectives center on sample return, Endurance's long traverse would enable a variety of additional transformative science investigations. |
|---|---|

Endurance's nearly 2,000-kilometer traverse is motivated by requiring samples from specific locations, sampling the geologic diversity of SPA. However, it is important to recognize that **a rover traverse of this magnitude would provide an unprecedented opportunity for a variety of in situ science investigations far exceeding the minimum required for selecting and contextualizing the samples.** There was a strong desire within the Endurance Science Workshop to capitalize on this opportunity. In this section, we provide three examples of such investigations. These examples are broad and non-exhaustive, and only serve to highlight the potential of the Endurance concept. These examples may, or may not, constitute essential geologic context as defined in **Finding #3**.

## VOLATILES

Endurance's baseline traverse would take it from the mid-latitudes (approximately 55°S) to the south pole (with the exact latitude dependent on the nature of the Endurance-Artemis rendezvous). This traverse, spanning approximately 35° in latitude (**Figure 10**), would provide an opportunity to characterize the extent and nature of volatiles (*e.g.*, hydrogen, water, hydroxyl) on the lunar surface along a continuous traverse from well-illuminated, volatile-depleted regions at mid-latitudes, to the pole and its prominent, volatile-rich permanently shadowed regions (PSRs). As Endurance traverses this region, it may pass by an increasing number of micro cold traps[38] (PSRs at the scale of centimeters to kilometers), transiently shadowed regions (TSRs), and potentially larger PSRs. Many mapped PSRs are near the baseline Endurance traverse, with the first (most northernmost) being located in the vicinity of Schrödinger basin (~72°S). Even outside of PSRs, volatiles have been observed in some low-latitude sunlit region[39], with abundances that vary as a function of the lunar day[40]. Endurance would operate over multiple lunar day/night cycles, potentially providing an opportunity to characterize volatiles as a function of time. Endurance's long-range traverse could provide important regional/global context for other missions to the lunar surface that focus on single locations (*e.g.*, CLPS missions), or explore a localized area (*e.g.*, VIPER[5], and early Artemis missions).

---

[38] Hayne et al. (2021), Nature Astronomy: https://www.nature.com/articles/s41550-020-1198-9.

[39] Honniball et al. (2021), Nature Astronomy: https://www.nature.com/articles/s41550-020-01222-x.

[40] Hendrix et al. (2019), Geophysical Research Letters: https://doi.org/10.1029/2018GL081821.

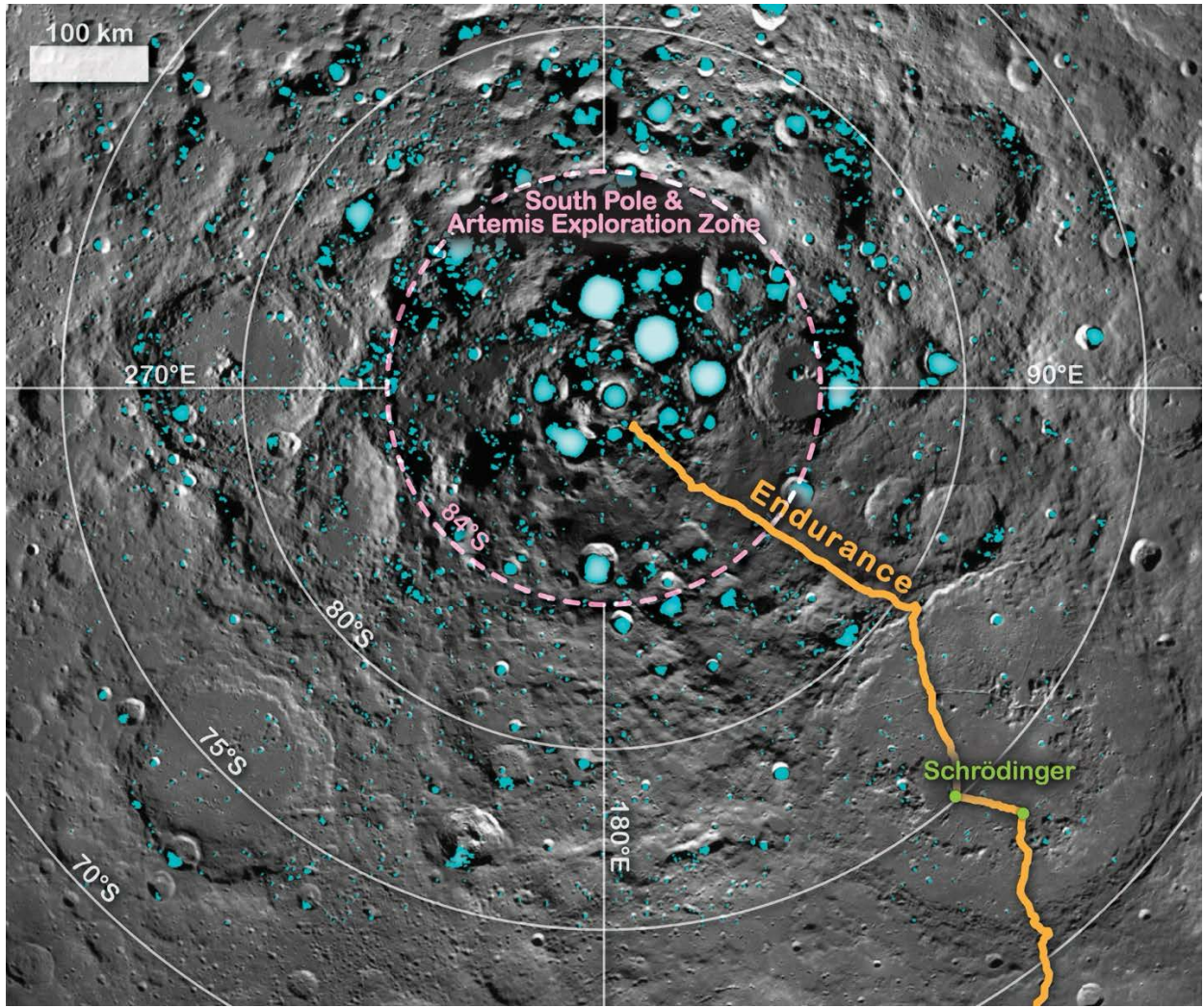

*Figure 10: Endurance's traverse in the context of the Moon's volatiles and permanently shadowed regions (PSRs).* PSRs greater than 1 km$^2$ are shown in light blue. Endurance's baseline traverse is shown in gold, with two visible sample sites in Schrödinger basin shown in green. The Artemis exploration zone (poleward of 84°S) is outlined in pink. Notice that Endurance begins to pass PSRs as far north as Schrödinger, and will pass many large PSRs towards the end of the mission near the south pole. The exact end of the Endurance traverse is yet to be determined, as it will require coordination with Artemis for the astronaut rendezvous and transfer of samples. Figure created with LROC Quickmap by James Tuttle Keane.

The baseline Endurance traverse would not drive through any PSRs as there was no requirement to do so, and to avoid any potential hazards. However, the baseline Endurance rover would almost certainly be capable of exploring PSRs, due in large part to its baseline power system (RTG). This capability was studied during the Decadal

Survey as part of the INSPIRE concept study[41] for a long-range, PSR exploring, Intrepid-derived rover that would explore the largest, oldest, and coldest PSRs at the Moon's south pole. There are many accessible PSRs near the current baseline traverse, with sufficiently benign slopes to allow Endurance to drive in.

Instruments onboard Endurance may be capable of characterizing volatiles. As described previously (**Section 3**), one of the ground rules for the Endurance concept study was that Endurance would baseline the same instrument suite as the Intrepid concept study. Amongst these instruments, only one—the Gamma Ray and Neutron Spectrometer (GRNS)—was levied with requirements for detecting hydrogen (at the 50-ppm level). GRNS is a notable example of a type of instrument that can do both volatiles, but also provide in situ measurements germane to Endurance's sample science and geologic context objectives (*e.g.*, measuring thorium abundance as a tracer for mantle material). There are countless other instruments and possible approaches for characterizing volatiles from a rover-based platform, including (but not limited to) spectrometers spanning from the far ultraviolet to the far infrared, electromagnetic sounding methods, thermal measurements, and more. However, while many different kinds of complementary measurements and investigations were presented at the workshop, it is important that the sample characterization, collection, contextualization, and return activities take precedence in an Endurance concept.

## GEOLOGY AND GEOPHYSICS

As Endurance traverses the South Pole–Aitken basin, it will visit (almost) countless diverse terrains, both young and old, shaped by a variety processes—from impact craters of all sizes to volcanic deposits of varying source and composition, and a myriad of types and ages of regolith. There is substantial opportunity for advancing our knowledge of fundamental geological and geophysical processes relevant to the Moon and other solar system worlds. While many of the Endurance mission science goals are centered around sample return analyses, other knowledge may be gained about the geology of South Pole–Aitken basin during the rover's traverse. This is important for multiple reasons. First, in situ geology and geophysics measurements provide critical geologic context for the samples collected by Endurance. Second, in situ measurements may enable some science outcomes, even if the samples are not returned to Earth for some unforeseeable reason.

As described previously (**Section 2**), one of the ground rules for the Endurance concept study was that it would utilize the same instrument suite as the Intrepid concept study (detailed in **Table 2**) which was developed as a Pre-Decadal Mission Concept Study[3]. The Endurance concept study recognized this as an important topic for further study. During the Endurance Science Workshop, participants discussed how a variety of instruments would be capable of addressing a range of open questions in lunar geology and geophysics. Themes included (but were not limited to):

- **Impacts:** Impacts are a fundamental planetary process that shape worlds across

---

[41] INSPIRE (In Situ Solar System Polar Ice Roving Explorer) concept study report (and all other Decadal Survey concept study reports): https://tinyurl.com/2p88fx4f.

the Solar System. Endurance would drive through (or around) almost innumerable impact craters ranging in size and age from the ancient SPA "mega basin," to impact basins (*e.g.*, Poincaré, Apollo, Schrödinger), and countless smaller impact craters that dominate the lunar farside. Endurance could provide an unparalleled opportunity to study impacts as a geologic process. In situ characterization of impact features along the traverse may also help contextualize Endurance's returned samples, and assist scientists in determining the provenance of rock clasts from different sites.

- **Volcanism:** SPA is host to a variety of volcanic landforms, ranging from mare deposits to pyroclastic deposits. SPA would provide an opportunity to probe these diverse volcanic landforms and understand what they tell us about the lunar interior, and how they differ from nearside volcanic deposits sampled by Apollo and other missions. In situ measurements of volcanic features along the traverse may also help scientists determine the provenance of volcanic rocks in Endurance's returned samples.

- **Tectonism:** There are a variety of tectonic landforms within SPA. Of particular interest are younger wrinkle ridges, which may even be tectonically active today[42]. Determining the level of present-day tectonic activity may inform our understanding of the Moon's thermal evolution, and inform hazard analyses for future explorers.

- **Regolith formation and evolution:** A 2,000-kilometer traverse would provide an opportunity to study the formation and modification of regolith which may inform our understanding of many other airless worlds. Traversing a wide range of latitudes over multiple day/night cycles may also inform space weathering and thermal breakdown of regolith, which may vary as a function of space, time, and terrain type.

- **Magnetic anomalies:** Endurance's baseline traverse may drive through magnetic anomalies. The nature of the Moon's magnetic anomalies is widely debated, as is the evolution of the lunar core dynamo. Surface measurements of the lunar magnetic field is particularly important, as orbital measurements are inherently limited since they are widely separate from the source. A 2,000-kilometer magnetic traverse would be unprecedented in planetary science, and more akin to terrestrial magnetic traverses which have been transformative for terrestrial geology[43]. In situ magnetic measurements are also highly complementary to magnetic measurements of returned samples.

- **Special environments, including permanently and transiently shadowed regions:** Endurance would traverse through (or near) many special environments on the Moon with unique thermal and illumination conditions. These locations could be important for understanding lunar volatiles (described in the previous section)

---

[42] Nypaver & Watters (2024), Lunar and Planetary Science Conference: https://www.hou.usra.edu/meetings/lpsc2024/pdf/1938.pdf.

[43] Hamoudi et al. (2011), Aeromagnetic and Marine Measurements, in "Geomagnetic Observations and Models": https://link.springer.com/book/10.1007/978-90-481-9858-0.

and how the surface interacts with the space environment.

Endurance is uniquely poised to make major advances in these themes, if Endurance is appropriately designed, instrumented, and operated.

There was substantial discussion at the Endurance Science Workshop about the myriad of measurement types that could do investigate the geology and chemistry across the diverse terrains of the lunar farside. For example, there was strong desire for continuous full-color, high-definition video and stereo imagery with a multi-spectral terrain camera. This could provide critical geologic context along the route (and have the added benefit of being an amazing public engagement tool). Hyperspectral instruments can provide more quantitative insight to the local chemistry. For example, visible (0.3–0.8 microns), near-infrared (0.7–4.0 microns), and intermediate infrared (4–8 microns) spectrometers could determine in situ mineral compositions, including olivine and pyroxene composition (*e.g.*, Mg#), or even detection of water and hydroxyl. Thermal infrared (4–100 microns) spectrometers can measure the surface temperature, thermal inertia, roughness, and abundance of rocks and soils—which can be important for geologic context and sample selection. Thermal infrared can also be useful for accurately measuring extremely cold temperatures, below 50 K, and understanding the thermal environment and cold-trapped volatiles. APXS and GRNS can provide elemental abundances. X-ray instruments are also powerful, including x-ray diffraction (XRD) which can provide mineral structure, x-ray fluorescence (XRF) can provide elemental compositional data, and x-ray micro computed tomography (XCT) can provide three-dimensional micromorphological analyses of samples. LIBS can also provide some chemical information from a distance. Two challenges for many in-situ measurements are dust coverings of rocks, and space weathering rinds, both of which can obscure the true chemistry of the bulk rock. There may be different ways of mitigating this based on the type of instrument and the desirable amount of complexity (*e.g.*, brushes, abrasion tools, etc.). Some measurements, like observations of rock texture, are less sensitive to these effects.

The long traverse of Endurance is a rare opportunity to integrate orbital and surface measurements to understand the structure of the Moon's crust. A sufficiently sensitive rover-based accelerometer—be it the rover's inertial measuring unit (IMU) or a dedicated gravimeter—would permit measurements of the lunar gravitational field along-traverse at a much higher spatial resolution than what was achieved by the Gravity Recovery and Interior Laboratory (GRAIL) mission. GRAIL provided a superb orbital dataset with spatial resolution of ~5 kilometers[44]. Depending on the instrument and concept of operations, a rover could make measurements at much finer spatial scale. Rover-based gravity measurements have been demonstrated by Mars rovers, including Curiosity at Gale Crater[45]. The resulting data could potentially resolve the fine-scale vertical structure of the Moon and enable imaging of the lunar crust, Mohorovičić discontinuity (the boundary between the crust and mantle), and mantle. In addition, gravity data could elucidate the subsurface structure and compositions of impact basins

---

[44] Goossens et al. (2019), Journal of Geophysical Research: Planets: https://doi.org/10.1029/2019JE006086.

[45] Lewis et al. (2019), Science: https://doi.org/10.1126/science.aat0738.

Final Report of the
Endurance Science Workshop 2023and craters, help map the megaregolith, and enable scouting of lava tubes which may serve as habitats in future lunar colonization.

Ground penetrating radar (GPR) may be a particularly compelling instrument for studying the lunar regolith, megaregolith, and shallow crustal structures. While the original Endurance concept study did not include GPR, GPR has flown on several planetary rovers in recent years, including the Radar Imager for Mars's Subsurface Experiment (RIMFAX) instrument on Perseverance, and the GPR onboard the Chang'e 3 and Chang'e 4 rovers (Yutu and Yutu 2, respectively). The Yutu rovers have demonstrated the ability to probe the lunar regolith, including mapping subsurface regolith and lava flows to several hundred meters depth [46,47]. The forthcoming Cooperative Autonomous Distributed Robotic Exploration (CADRE) rovers will also carrying multi-static GPR on a forthcoming CLPS mission. Thus far, GPR has only been deployed on relatively short traverses (10s of meters to a few kilometers). Endurance's 2,000-kilometer-long traverse would provide an opportunity to probe subsurface structure across a major transect of SPA and a variety of terrains therein. Magnetotellurics and other electromagnetic sounding approaches could also contribute to developing an understanding of the vertical layering of the crust, including composition and structure.

Because the Endurance traverse would drive over regions of the lunar crust with vastly differing ages, magnetometer measurements of remanent crustal fields acquired during the Endurance traverse may provide a window into the evolution of the ancient lunar dynamo. Electron reflectometry indicates that regions along the proposed Endurance A-traverse that appear relatively unmagnetized from orbital magnetometer measurements may in fact contain significant crustal remanence[48]. The spatial resolution enabled by a rover surface traverse would be a transformative improvement over what is achievable with orbital spacecraft magnetometer and electron magnetometer data and will provide ground truth for prior crustal field properties inferred from altitude. Rover magnetometer measurements would also elucidate the geometries of magnetic source bodies (especially if gravity and magnetic data are utilized in combination), test the contrasting hypotheses of endogenic vs. exogenic origins for strongly magnetic material in the lunar crust, and explore linkages between crustal fields and space weathering. To ensure cleanliness of rover magnetometer measurements, it is recommended that a boom be used to keep the magnetic field sensor away from the rover and that engineers take steps to produce magnetically quiet rover subsystems and implement other field mitigation measures (*e.g.*, magnetic shielding, signal subtraction).

In situ geophysics has long been a goal for the planetary science community, with a Lunar Geophysical Network (LGN) being consistently recognized as a high priority mission concept for several Decadal Surveys. There was considerable discussion at the Endurance Science Workshop about how Endurance could contribute to the goals of LGN. For example, it is plausible to imagine Endurance carrying and then deploying

---

[46] Xiao et al. (2015), Science: https://doi.org/10.1126/science.1259866.

[47] Lai et al. (2020), Nature Communications: https://doi.org/10.1038/s41467-020-17262-w.

[48] Szabo et al. (2023), Endurance Science Workshop 2023: https://www.hou.usra.edu/meetings/endurance2023/pdf/3021.pdf.

Page 39

standalone scientific payloads including seismometers. The large payload and mobility capabilities of Endurance could enable deployment of geophysical packages in optimal locations, or even deployment of novel payloads—like laying of fiber optic cable to enable Distributed Acoustic Seismology (DAS). Even without dedicated seismometers, the rover's IMU may permit limited seismic measurements if coupled with some form of active seismic source (*e.g.*, rover-launched projectiles or drill[49]). An IMU-based approach was proposed for the VIPER[5] mission, which may provide insight about near-surface regolith properties including porosity and structure which may be helpful for identifying ice layers. In addition to seismology, LGN also has goals for understanding the lunar heat flow. In situ heat flow measurements—either with direct thermal probes or long-wavelength (infrared, microwave, radio) radiometers—could provide constraints on local geothermal gradients along the traverse.

## SOLAR AND SPACE PHYSICS

As Endurance traverses approximately 2,000 kilometers from the mid-latitudes (~55°S) to the lunar south pole (~90°S), it would explore a variety of environments relevant to understanding how the space environment interacts with the lunar surface (**Figure 11**). The rover would traverse through a range of different geologic terrains (*e.g.*, basalt-rich volcanic deposits, anorthositic highlands, pyroclastics, etc.), through a range of differently magnetized terrains, and through a variety of illumination conditions—all while the Moon oscillates through solar wind and Earth space environment (terrestrial foreshock, magnetosheath, magnetotail) over several years (baseline mission duration is 4 years including margin, corresponding to approximately 49 lunar days). This combination of environments may make Endurance an intriguing in situ laboratory for addressing heliophysics science questions[50,51].

Endurance's returned samples may also provide an important record of solar activity[52]. As solar wind, cosmic rays, and meteorites bombard the surface of the Moon, they alter the chemical, isotopic, and petrologic makeup of surface material. Owing to the Moon's lack of atmosphere and relative paucity of geologic activity, these alteration signatures may be preserved through time—plausibly providing a record of space weather (including the flux of solar energetic particles, cosmic rays, gamma ray bursts, and supernova) going back billions of years. By appropriately collecting regolith samples,

---

[49] Lewis et al. (2022), Fall Meeting of the American Geophysical Union: https://agu.confex.com/agu/fm22/meetingapp.cgi/Paper/1141984.

[50] Halekas et al. (2022), white paper for the Decadal Survey for Solar and Space Physics (Heliophysics) 2024-2033: http://surveygizmoresponseuploads.s3.amazonaws.com/fileuploads%2F623127%2F6920789%2F121-878c61ce386b12678b157ef920ab46c7_HalekasJasperS.pdf.

[51] Heliophysics Science and the Moon: Potential Solar and Space Physics Science for Lunar Exploration (2007), NASA report: https://science.nasa.gov/science-red/s3fs-public/atoms/files/Final_508Compliant_MoonRpt.pdf.

[52] Saxena et al. (2022), white paper for the Decadal Survey for Solar and Space Physics (Heliophysics) 2024-2033: https://arxiv.org/pdf/2208.13307.pdf.

Endurance could address this complementary aspect of Solar System history.

**The possible implications for solar and space physics demonstrates that Endurance has the potential for impact across multiple divisions within NASA's Science Mission Directorate.**

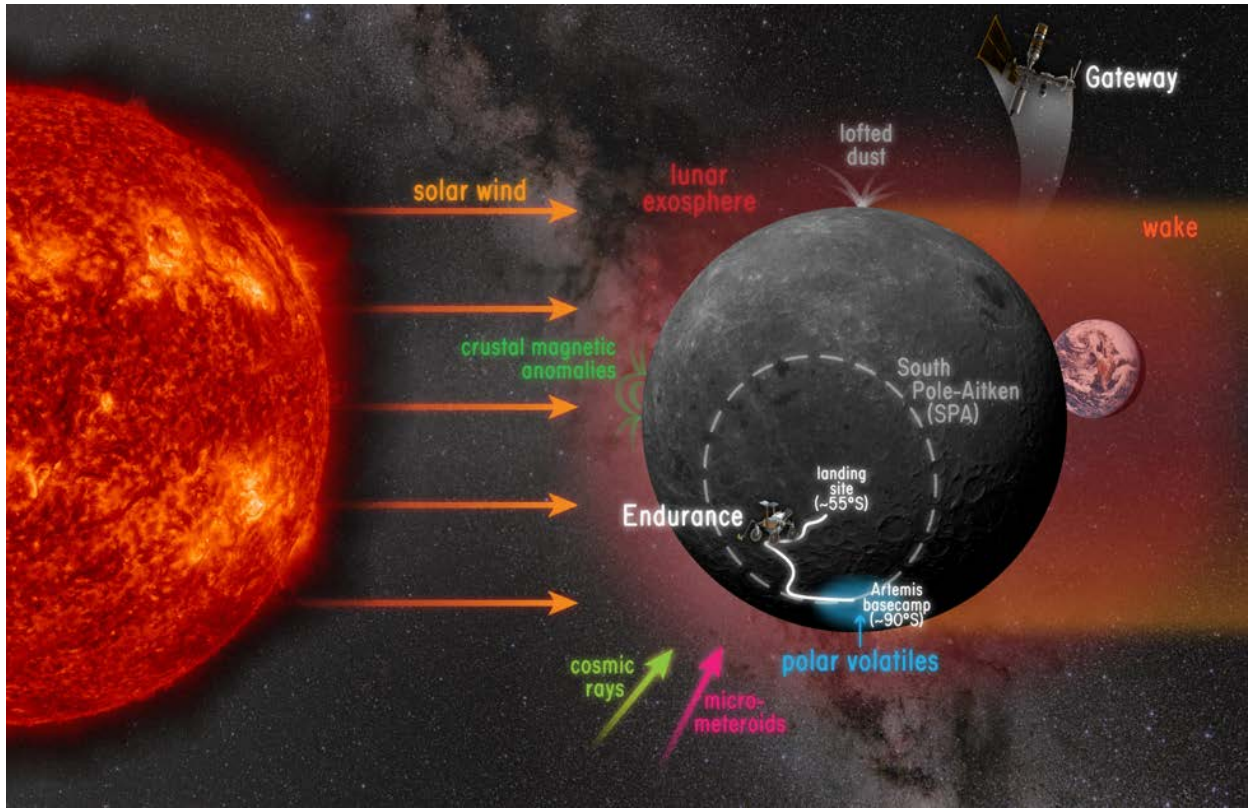

*Figure 11: Endurance traverse in the context of the Moon's space environment. Endurance has the potential to make insightful observations over a range of lunar latitudes (from 55°S to 90°S), times of day, and lunar terrain types (e.g., magnetic anomalies) that may enable investigations about space weathering and interactions between the space environment and the surface of the Moon. Graphic by James Tuttle Keane.*

| FINDING #5 | Endurance is an ambitious mission that would be enabled and enhanced by investing in developing key technologies now. |
|---|---|

Endurance represents a new way to explore planetary surfaces. A long-lived, long-distance, highly autonomous rover would be a dramatic shift in robotic exploration tactics and capabilities—but is necessary to enable accomplishing Endurance's ambitious science objectives. This new approach is possible because it leverages substantial technology developments in recent years, including some currently being implemented on the Curiosity and Perseverance rovers on Mars. While the Endurance Science Workshop was not explicitly focused on the technologies underpinning Endurance, there was substantial discussion about how new technologies could enable or enhance the science return from this mission. In this section, we summarize the key points from the Workshop. A central theme from this discussion is that **it is imperative that the SDT quantify science requirements that can inform near-term technology investments for Endurance.**

## DRIVING AUTONOMY

Endurance's 2,000-kilometer traverse is unprecedented in planetary exploration (**Figure 12**). To accomplish this drive in a reasonable mission duration (baseline mission duration is four years, *i.e.*, forty-nine lunar days), Endurance must be capable of driving autonomously and reliably during both the lunar day and night with less frequent ground oversight than current Mars rovers. Ground-in-the-loop driving—either with direct human control (*i.e.*, "joy-sticking") or with supervised ground-based computing—is not capable of meeting the necessary traverse speeds (of ~1 kilometer per hour) given the expected operational constraints, such as the limited availability of communications relays.

While driving autonomy was not discussed substantially at the Endurance Science Workshop, it was recognized that this is a critical technology that enables the Endurance concept. This autonomous driving capability have been detailed in the Endurance concept study report[2] and other reports[53].

---

[53] Baker et al. (2024), IEEE Aerospace Conference: https://doi.org/10.1109/AERO58975.2024.10520939.

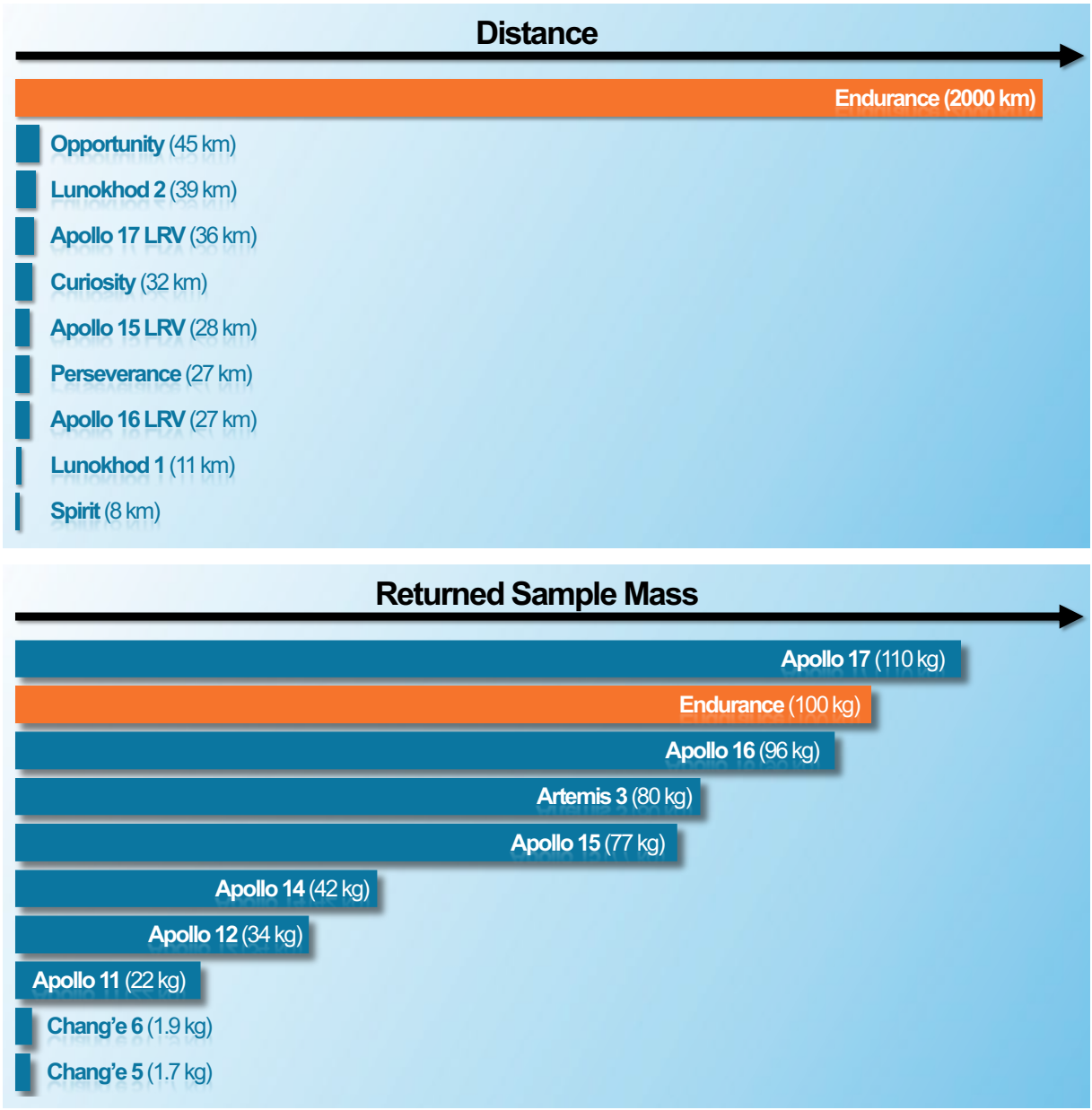

*Figure 12: How does Endurance's traverse distance (top) and returned sample mass (bottom) compare with other missions?* Values for future missions, including Endurance and Artemis III are subject to change: Endurance's baseline distances and sample masses are shown; Artemis III returned sample mass is based on the goal mass for returned payload exclusive of containers as defined in the Artemis III Science Definition Team report[21]. Missions with drive distances less than 5 kilometers, and sample return masses less than 1 kg are not shown.

## FARSIDE COMMUNICATIONS AND DATA VOLUME

As Endurance traverses the farside of the Moon, it has the potential to acquire unprecedented scientific data, both in quantity and quality. This in situ data would need to be relayed back to Earth via a communications relay satellite(s). At present, there are multiple relay satellite concepts in development that would meet the baseline requirements of Endurance as defined in the concept study report[54]. However, the ability of communications infrastructure to meet Endurance's requirements invariably depends on the final choice of instrument suite and concept of operations. Many instruments could produce large data volumes. A notable example includes imaging spectrometers[55,56] which would produce spectra for every pixel in an image (a data cube) that would far exceed the data volume for point spectrometers—like the point spectrometer baselined in the Intrepid and Endurance concept studies. Another notable example is XCT instruments that could characterize the internal 3D structure of samples. While these instruments would dramatically increase data volume demands, they may substantially enhance the science return, including providing far more geologic context (like mapping mineralogy around the rover) that would directly support Endurance's sample return objectives.

Given the potential need for high data volumes from Endurance, it is important that NASA and the broader lunar community continue to consider the importance of high data volumes when developing lunar relay infrastructure. As noted by a participant at the Endurance Science Workshop, "Endurance is an Earth-orbiting mission"[57] and the community should aspire for data rates commensurate with other Earth-orbiting missions. High data volumes from the Moon will likely be required for human operations as part of Artemis, and these capabilities should be extended to other scientific missions.

The need for high data rates and volumes from the Moon was highlighted in the recent LEAG Continuous Lunar Orbital Capabilities Specific Action Team (CLOC-SAT) report[58] that identified the science and exploration needs for future lunar orbiters:

> *"Overarching Finding 8: Lunar surface and orbital data acquisition and relay are not severely constrained by transmission distance; therefore, robust data throughput (data rate) must become standard at the Moon in order to realize the potential of next-generation instrumentation. A holistic strategy for processing and archiving these large data sets is also needed."* (CLOC-SAT report, page 68)

---

[54] Endurance baselined the European Space Agency Lunar Pathfinder communications relay, with 3 to 8 hours per day of telecom passes (depending on mission phase) with a transmit rate of 8 megabits per second. The total returned data volume over the four-year baseline mission was approximately 5.5 gigabytes, which provided about 60% margin on the required data volume.

[55] Kremer et al. (2023), Endurance Science Workshop: https://www.hou.usra.edu/meetings/endurance2023/pdf/3017.pdf.

[56] Fraeman et al. (2023), Endurance Science Workshop: https://www.hou.usra.edu/meetings/endurance2023/pdf/3015.pdf.

[57] Quote from Carlé Pieters, Endurance Science Workshop (2023).

[58] LEAG CLOC-SAT Report (2023): https://www.lpi.usra.edu/leag/reports/CLOC-SAT_Report.pdf.

While high data rate relay capabilities are desired, approaches for decreasing Endurance's data volume should also be considered. Onboard data analysis and science autonomy may provide an additional method by which large datasets could be reduced in the event that the data volume returned to Earth is limited (see the next section). Additionally, since Endurance relies on Artemis astronauts to ultimately retrieve Endurance's sample cache, one could imagine that astronauts could also retrieve other pieces of hardware—such data storage—for download or return to Earth. While this would levy additional requirements on the mission (a data storage device that is interoperable with Artemis), this may be an intriguing option should Endurance be capable of collecting more data than it can transmit back to Earth.

## SCIENCE AUTONOMY

Onboard science autonomy describes the ability of a spacecraft to autonomously analyze raw scientific data, make real-time decisions based on those analyses, and summarize or prioritize data for transmission back to Earth. Science autonomy has been infused in active missions to varying degrees. For example, the Perseverance rover on Mars utilizes the AEGIS[59] (Autonomous Exploration for Gathering Increased Science) system to analyze rover camera images, identify and prioritize targets based on science criteria (*e.g.*, rock size, shape, brightness), and then acquire follow-up, high-resolution data using other instruments including LIBS and other spectroscopic data. The Decadal Survey identified science autonomy as a key technology area needing development in the coming decade and beyond (see page 544 of the Decadal Survey[1]).

It is important to distinguish science autonomy from driving autonomy. We define driving autonomy as the ability of the rover to autonomously navigate, drive, and operate on the surface of the Moon with little/no human intervention. Driving autonomy is a requirement for Endurance (as discussed in a previous section), but was not a focus of the Endurance Science Workshop. Science autonomy is not currently a requirement for Endurance. Nonetheless, the Endurance Science Workshop did identify science autonomy as a particularly compelling technology that may substantially enhance the Endurance concept. In this section, we describe several possible examples of science autonomy that could be applicable for Endurance.

### *EXAMPLE SCIENCE AUTONOMY USE CASE: ANTICIPATING SURPRISES*

While Endurance's concept of operations relies on a great deal of pre-planning, our past experiences with lunar and planetary exploration have demonstrated that we should always be prepared for surprises. Perhaps the best, most relevant example was the discovery of the "orange soil" by Jack Schmitt and Eugene Cernan on Apollo 17 (**Figure 13**). This material—which was later realized to represent volcanic glass—was not anticipated from orbital or telescopic observations, and only discovered by an observant astronaut walking across (and stirring up) the lunar surface. One can only imagine the surprises Endurance may find along its 2,000-kilometer traverse (roughly 56-times further

---

[59] Verma et al. (2023), Science Robotics: https://www.science.org/doi/full/10.1126/scirobotics.adi3099.

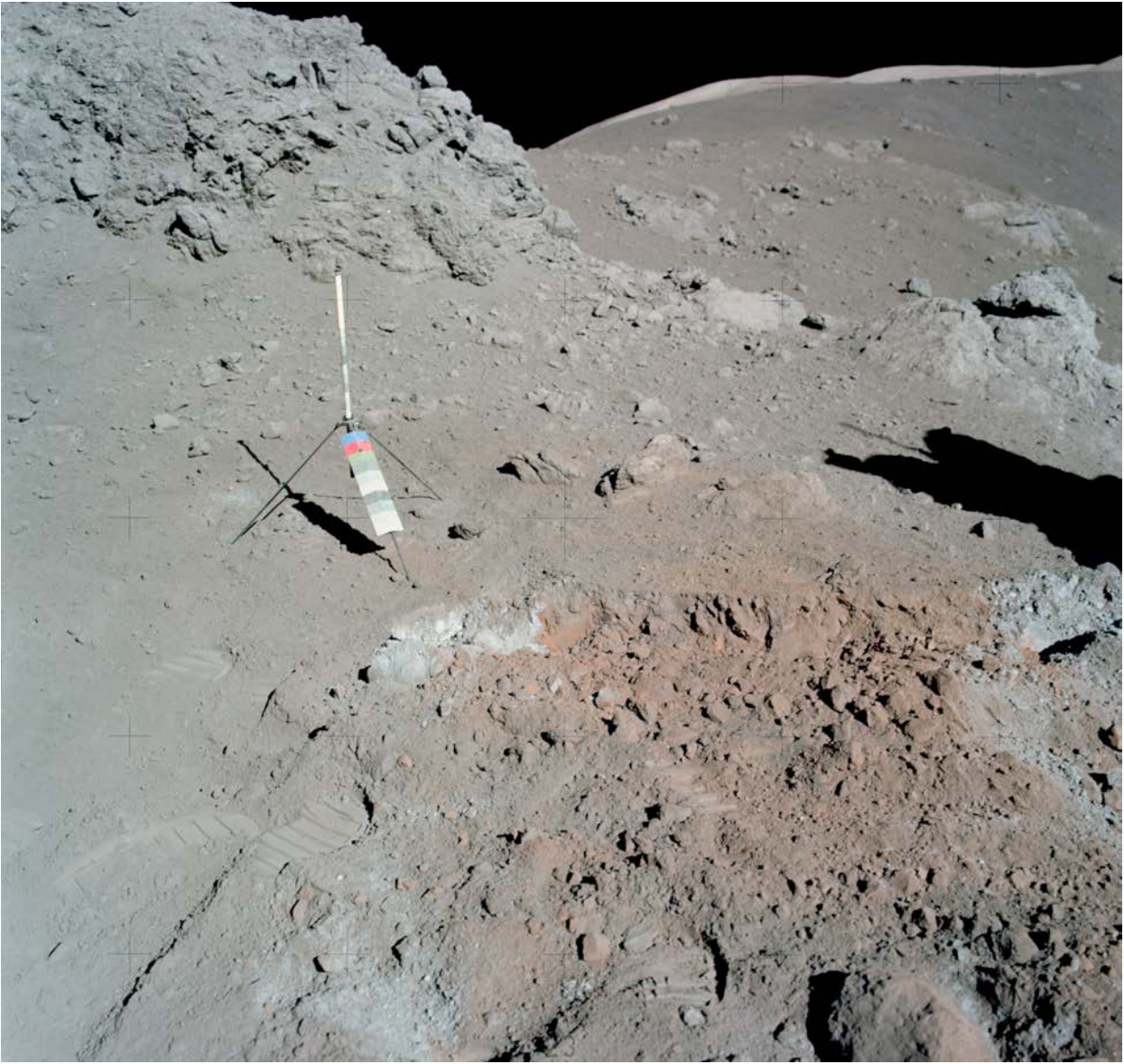

**Figure 13: An unexpected surprise on a drive: the orange soil discovered by Jack Schmitt on Apollo 17.** *It is possible that Endurance would make similar unexpected discoveries while traversing an unprecedented distance across the farside of the Moon. Science autonomy may enable Endurance to make those surprising discoveries. Image credit: NASA AS17-137-20990.*

than the distance covered by the Apollo 17 astronauts on their lunar roving vehicle).

Onboard science autonomy may enable Endurance to both identify, and be responsive to, unexpected discoveries with reduced impact to the overall mission schedule. For example, consider if Endurance acquired full-color images at a more frequent cadence to search for features of interest (perhaps every few hundred meters, instead of every few kilometers). With traditional rover operations, those images would need to be transmitted to Earth (which could take between seconds to hours, depending on the communications relay infrastructure), analyzed by scientists, and then the ground team would need to make decisions about whether to stop, go-back, or proceed on. This can take a considerable amount of time—which ultimately slows the rover. Instead, imagine if Endurance possessed the capability to both take data at a more frequent cadence and autonomously analyze the data onboard, Endurance could make its own decisions about when to stop, send data home, or even take additional measurements (*e.g.*, take more images or spectra). Science autonomy also need not be limited to analyses of imagery data; for example, one could imagine real time analyses of magnetometer data to identify (and perhaps map and localize) small-scale magnetic anomalies—like those thought to be present due to magmatic intrusions or the metallic remnants of impactors.

In the baseline concept, Endurance did not prioritize exploration in the sense of searching for anomalous terrains or unexpected discoveries. Rather, Endurance explicitly prioritized getting from one sample site to the next as quickly as possible and accomplishing its science objectives (**Table 1**) within the allocated schedule and budgetary constraints. While this did include frequent science stops along the traverse (*e.g.*, "interval stops" every 2 kilometers where Endurance would acquire full color panoramas, and utilize its full remote sensing payload) to provide geologic context for the samples, this did not include allocating time for detours. I**t would be incumbent on the future Endurance SDT (Box 1) to outline what would be acceptable reasons for deviating from the pre-planned traverse (if any).**

### *EXAMPLE SCIENCE AUTONOMY USE CASE: REDUCING DATA VOLUME*

As noted previously, many potential science instruments produce large data volumes. Imaging spectrometers are a classic example where the instrument can produce "data cubes" with spectral information for every pixel in a 2D image. Imaging spectrometers have a long heritage for operations at the Moon, including $M^3$ aboard Chandrayaan-1 and the $HVM^3$ aboard the forthcoming Lunar Trailblazer mission[60]. **Figure 14** shows an example of data from a terrestrial imaging spectrometer and how it can be used to image a complex scene (an impact structure in Canada) and, by application of spectral models, be used to produce derived datasets including lithology maps.

One approach to accommodate high data-volume instruments in data-limited missions is to have the spacecraft autonomously analyze the data and only transmit back derived data products. In the imaging spectrometer example shown in **Figure 14**, one could imagine that instead of transmitting back the entire data cube, you transmit just the lithology map. Onboard data analysis is a developing technology in planetary science and may be worth consideration for Endurance, depending on the science required levied by the forthcoming SDT (**Box 1**).

---

[60] Ehlmann et al. (2022), IEEE Aerospace Conference: https://doi.org/10.1109/AERO53065.2022.9843663.

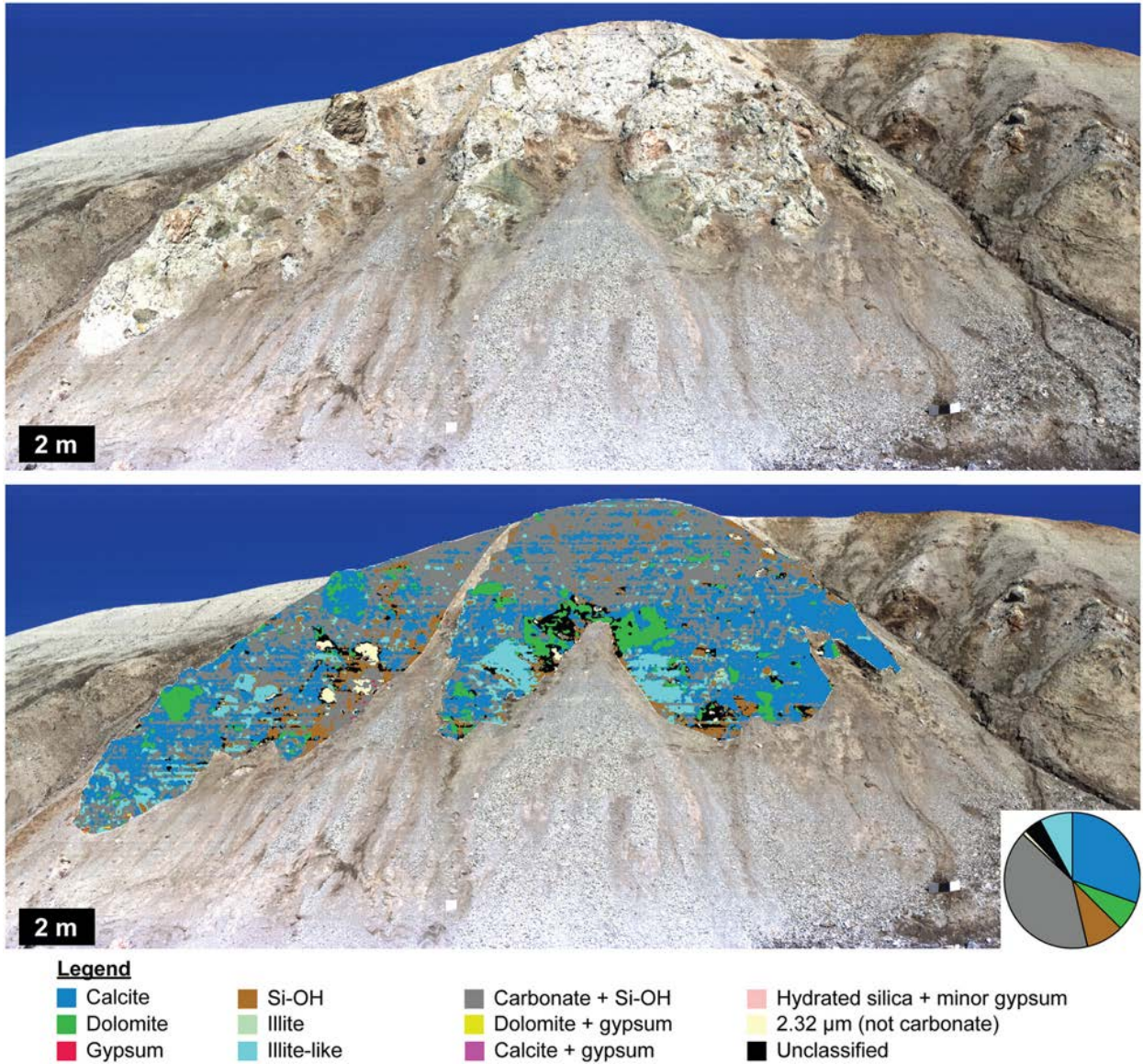

*Figure 14: An example of imaging spectroscopy methods applied in a terrestrial setting: the Rhinoceros Creek outcrop at the Haughton impact structure in Canada. The top image shows a color composite, while the bottom image shows a lithologic map. The lithologic map was constructed by analysis of data from a visible/infrared imaging spectrometer (0.4–2.6 microns). This image demonstrates the rich diversity of lithologies that could be present in well-mixed rocks (e.g., impact breccias, regolith), and how computer algorithms could be used to potentially identify target lithologies. Image courtesy of Rebecca Greenberger and Bethany Ehlmann[61]. This outcrop may not be representative of lunar terrains, and is only shown here for illustrative purposes.*

---

[61] Greenberger et al. (2020), Journal of Geophysical Research: Planets: https://doi.org/10.1029/2019JE006218.

### *EXAMPLE SCIENCE AUTONOMY USE CASE: SAMPLING*

The lunar surface that Endurance would traverse is covered by a thick layer of regolith—unconsolidated, fragmented, loose, heterogeneous material—created by eons of bombardment from impactors (ranging in scale from micrometeorites to catastrophic, basin-forming asteroids and comets). This material is further mixed by the impact process, which moves material both vertically and laterally (so-called "impact gardening"). When exposed to space, regolith is also irradiated by charged particles from the Sun and other stars (so-called "space weathering"). The mixing of lunar material is even present at the scale of individual rocks, particularly impact breccias where regolith is lithified together. This makes the lunar surface fundamentally different from many other planetary bodies, like the Earth and Mars. There is effectively little/no bedrock on the Moon.

The mixing of lunar regolith presents an interesting conundrum for lunar sample return missions like Endurance. Endurance's science objectives (**Table 1**) require collecting samples of a specific lithology and size. Current (and near-future) remote sensing datasets enable us to identify specific locations on the Moon where we anticipate finding such samples. Depending on the scientific investigation in question, these sites can be specified ahead of time with ~1-meter precision when only visible imagery is required (such as finding a geologic contact), or ~100-meter precision when depending on mineralogy maps like those developed by the $M^3$ on Chandrayaan-1 and $HVM^3$ on the forthcoming Lunar Trailblazer mission. However, Endurance may ultimately need to use its in-situ instruments to identify material that meets its sample return requirements—whether those requirements are driven by science (*e.g.,* the need to select grains of a certain size and composition for geochronology investigations) and/or driven by technical capabilities and limitations (*e.g.,* the design of the sampling system). In the original Endurance baseline concept, it was envisioned that Endurance would utilize its visible cameras and infrared point spectrometer to collect data so that scientists on the ground could identify material that met sample size and lithology requirements[62], and that the mixing of lunar regolith would be mitigated by collecting an excessive volume of material (up to 8 kg per sample site) ensuring that the sample requirements were met by shear statistics. While this baseline concept of operations was sufficient for the concept study, discussion at the Endurance Science Workshop highlighted that the lunar regolith may not be amenable to this approach. Having scientists on the ground be responsible for analyzing data and identifying targets may be required, even though it would require substantial time, ultimately slowing the mission.

Having the rover autonomously perform some aspects of sample identification, characterization, and collection may enable Endurance to find more optimal samples in a shorter amount of time. Perhaps the most benign example would be if Endurance were able to use its cameras to identify patches of regolith with an optimal size-frequency distribution that met both science requirements and technical limitations for the sampling system. For example, imagine Endurance autonomously identifying patches of regolith with regolith particles near the sample size requirement (0.5–2.0 cm), and devoid of larger rocks that could create hazards for rover driving, arm placement, or sample system

---

[62] This is described in Appendix B of the Endurance concept study report[2].

Final Report of the
Endurance Science Workshop 2023clogging. A more exciting example for sample-related autonomy would be if Endurance used its spectral instruments to identify target lithologies for sampling. This type of real-time spectral analysis is now commonly done in terrestrial settings, as shown in **Figure 14**. While exciting, there was considerable debate about whether this more sophisticated approach would be appropriate for the Moon. We expect the impact gardened lunar regolith to be extremely well-mixed. Using hyperspectral instruments to pick specific rocks would imply the need for Endurance pick up specific rocks—which would dramatically impact the complexity of the sample acquisition system and operations. **It is important for the future SDT (Box 1) to carefully define what Endurance's true science requirements are, and what observations are necessary to select samples. There is a likely trade between the mass of the samples, and the complexity of the sample acquisition system and operations: as the mass allotted for samples increases, the need for contextualizing information may decrease.**

## TESTING

Endurance will face many testing challenges when preparing for operating in and sampling the lunar environment. A common refrain from the Endurance Science Workshop was "Be diabolical about testing"[63]. Our Mars experience has shown that testing in the right environment, with the right simulants are critical. In many ways, the Moon is an extreme environment—with enormous temperature swing (>350 K in daytime on the northern part of Endurance's traverse, <40 K in permanent shadows on the southern part of the traverse) in vacuum, and extremely angular and abrasive regolith that is challenging to replicate on the lab. It is important to challenge the system, and "be more creative about the ways that the Moon can mess you up"[63]. There have been substantial developments in both simulants, test environments, and laboratories across the country in the past several years that could be leveraged for Endurance. Examples include the Simulated Lunar OPErations (SLOPE) lab at NASA's Glenn Research Center[64] and the Lunar Lab and Regolith Testbeds at NASA's Ames Research Center[65].

---

[63] Quotes from Lori Shiraishi, Endurance Science Workshop (2023).

[64] NASA Glenn Research Center, Planetary Exploration Test Facilities: https://www1.grc.nasa.gov/facilities/slope-lab/.

[65] NASA Ames Research Center, Lunar Lab and Regolith Testbed: https://sservi.nasa.gov/testbed/.

Page 50

| FINDING #6 | Endurance should strive to include more diverse perspectives in its formulation—particularly from early-career scientists and engineers who will ultimately operate the rover and analyze the samples. |
|---|---|

As both a long-lived rover mission and a sample return mission, Endurance is poised to be one of the most impactful science investigations done on the Moon in the coming decades. A substantial portion of the lunar and planetary science community may participate in the mission in some form or another—be it from mission formulation, development and operations of instruments, planning the traverse, selecting samples, and analyzing them for the decades to come. In order to ensure the highest quality of science on the mission, it is crucial to be inclusive of diverse perspectives in the formulation and implementation of Endurance. Multiple studies have demonstrated that diversity is beneficial to the creativity, innovation, and scientific impact of teams[66,67,68,69]. Diversity and inclusiveness are also core values of NASA[70,71]. Despite the importance, demographics show that the planetary science workforce is not representative of the national population, and that many groups are underrepresented particularly in NASA planetary science missions[72,73,74,75].

---

[66] Hong and Page (2004), Proceedings of the National Academy of Sciences of the United States of America: https://doi.org/10.1073/pnas.0403723101.

[67] Campbell et al. (2013), PLOS One: https://doi.org/10.1371/journal.pone.0079147.

[68] Freeman & Huang (2014), Nature: https://doi.org/10.1038/513305a.

[69] Freeman & Huang (2014), National Bureau of Economic Research: https://www.nber.org/papers/w19905.

[70] NASA Science 2020-2024: A Vision for Scientific Excellence (2023): https://smd-cms.nasa.gov/wp-content/uploads/2023/09/2020-2024-nasa-science-plan-yr-23-update-final.pdf.

[71] NASA Strategic Plan (2022): https://smd-cms.nasa.gov/wp-content/uploads/2023/04/fy_22_strategic_plan-1.pdf.

[72] Porter et al. (2020), 2020 Survey of the Planetary Science Workforce by the Division for Planetary Sciences of the American Astronomical Society: https://dps.aas.org/dps-2020-membership-survey-results/.

[73] Rivera-Valentín et al. (2021), Bulletin of the American Astronomical Society: https://doi.org/10.3847/25c2cfeb.968ed505.

[74] Rathbun et al. (2021), Bulletin of the American Astronomical Society: https://doi.org/10.3847/25c2cfeb.da96f3af.

[75] National Academies of Science, Engineering, and Medicine: Advancing Diversity, Equity, Inclusion, and Accessibility in the Leadership of Competed Space Missions (2022): https://nap.nationalacademies.org/catalog/26385/advancing-diversity-equity-inclusion-and-accessibility-in-the-leadership-of-competed-space-missions.

Final Report of the
Endurance Science Workshop 2023The structure of the Endurance team should be considered from the earliest development stages to ensure the breadth of the lunar and planetary science community has the opportunity to participate. The lunar and planetary science community should also be prepared to reach beyond our community and include team members beyond planetary science including terrestrial geology, chemistry, robotic systems, and other science and engineering disciplines, as needed.

Importantly, it is not necessary to "reinvent the wheel" with regards to how to incorporate inclusion and diversity into planetary missions. A multitude of organizations have identified specific recommendations, ranging from the National Academies[1,75] to more organic organizations within/between the different NASA assessment and analysis groups[76]. NASA has developed its own programs, like the Here to Observe (H2O) program[77], and many NASA missions have their own initiatives[78]. One of the easiest methods for Endurance to engage with early-career community is via the Next Generation Lunar Scientists and Engineers (NGLSE or "NextGen") group[79]. NextGen provides professional development for early-career members of the lunar community.

---

[76] The Equity, Diversity, Inclusion, and Accessibility (EDIA) Working Group (WG): https://www.lpi.usra.edu/idea/working-group/.

[77] NASA Here to Observe (H2O) program: https://science.nasa.gov/planetary-science/programs/here-to-observe-h2o/

[78] Dragonfly Student & Early Career Investigator Program: https://dragonfly.jhuapl.edu/Student-Opportunities/.

[79] Next Generation Lunar Scientists and Engineers (NGLSE or "NextGen") group: https://www.nextgen-lunar.space/.

Page 52

# 5   ACKNOWLEDGMENTS

We would like to thank the participants of the Endurance Science Workshop—especially those who gave presentations, participated in panels, and moderated sessions. This workshop was only possible because of their scientific, technical, and programmatic contributions.

We thank the JPL and Caltech support team who provided the venue, audio/visual support, registration and other logistical support. This was particularly challenging given challenging given the hybrid nature of this workshop. Special thanks to John Baker, Martha Avina, Timothy Brice, Julie Castillo, Jann Overholt, and John Ziats.

We thank the Lunar and Planetary Institute (LPI) and Universities Space Research Association (USRA) support team who managed the Endurance Science Workshop website, registration, communications, and advertisement. Special thanks to Jamie Shumbera, and Monica George.

# 6 ACRONYMS

| | |
|---|---|
| ADS | Astrophysical Data System |
| AEGIS | Autonomous Exploration for Gathering Increased Science |
| AGU | American Geophysical Union |
| ANGSA | Apollo Next Generation Sample Analysis Program |
| APXS | Alpha Particle X-Ray Spectroscopy |
| ARMAS | Automated Radiation Measurements for Aerospace Safety |
| CADRE | Cooperative Autonomous Distributed Robotic Exploration |
| Caltech | California Institute of Technology |
| CAPS | Committee on Astrobiology and Planetary Science |
| CLOC-SAT | Continuous Lunar Orbital Capabilities Specific Action Team |
| CLPS | Commercial Lunar Payload Service |
| DAS | Distributed Acoustic Seismology |
| ESA | Electrostatic Analyzer |
| ESSIO | Exploration Science Strategy and Integration Office |
| FOV | Field of View |
| GPR | Ground Penetrating Radar |
| GRAIL | Gravity Recovery and Interior Laboratory |
| GRNS | Gamma Ray and Neutron Spectrometer |
| H2O | Here to Observe |
| HGA | High-Gain Antenna |
| HLI | Hand Lens Imager |
| HVM$^3$ | High-resolution Volatiles and Minerals Moon Mapper |
| IEEE | Institute of Electrical and Electronics Engineers |
| IFOV | Instantaneous Field of View |
| IMU | Inertial Measurement Unit |
| INSPIRE | In Situ Solar System Polar Ice Roving Explorer |
| JPL | Jet Propulsion Laboratory |
| LDEP | Lunar Discovery and Exploration Program |
| LEAG | Lunar Exploration Analysis Group |
| LED | Light Emitting Diode |
| LGA | Low-Gain Antenna |
| LGN | Lunar Geophysical Network |
| LIBS | Laser Induced Breakdown Spectroscopy |
| LOLA | Lunar Orbiter Laser Altimeter |
| LPI | Lunar and Planetary Institute |
| LPSC | Lunar and Planetary Science Conference |
| LRO | Lunar Reconnaissance Orbiter |
| LROC | Lunar Reconnaissance Orbiter Camera |
| LRV | Lunar Roving Vehicle |
| $M^3$ | Moon Mineralogy Mapper |
| NASA | National Aeronautics and Space Administration |
| NGLSE | Next Generation Lunar Scientists and Engineers (also known as "NextGen") |
| OSIRIS-REx | Origins, Spectral Interpretation, Resource Identification, and Security – Regolith Explorer |
| PLOS | Public Library of Science |
| PS | Point Spectrometer |
| PSD | Planetary Science Division |
| PSR | Permanently Shadowed Region |
| RGB | Red, Green, Blue |
| RIMFAX | Radar Imager for Mars's Subsurface Experiment |
| RTG | Radioisotope Thermoelectric Generator |
| SDT | Science Definition Team |
| SLOPE | Simulated Lunar OPErations |
| SMD | Science Mission Directorate |
| SPA | South Pole–Aitken |
| SPACA | South Pole–Aitken Compositional Anomaly |
| SPARX | South Pole–Aitken basin sample Return and eXploration |
| TSR | Transiently Shadowed Region |
| USRA | Universities Space Research Association |
| VIPER | Volatiles Investigating Polar Exploration Rover |
| WAC | Wide Angle Camera |
| XCT | X-Ray Computed Tomography |
| XRD | X-Ray Diffraction |
| XRF | X-Ray Fluorescence |

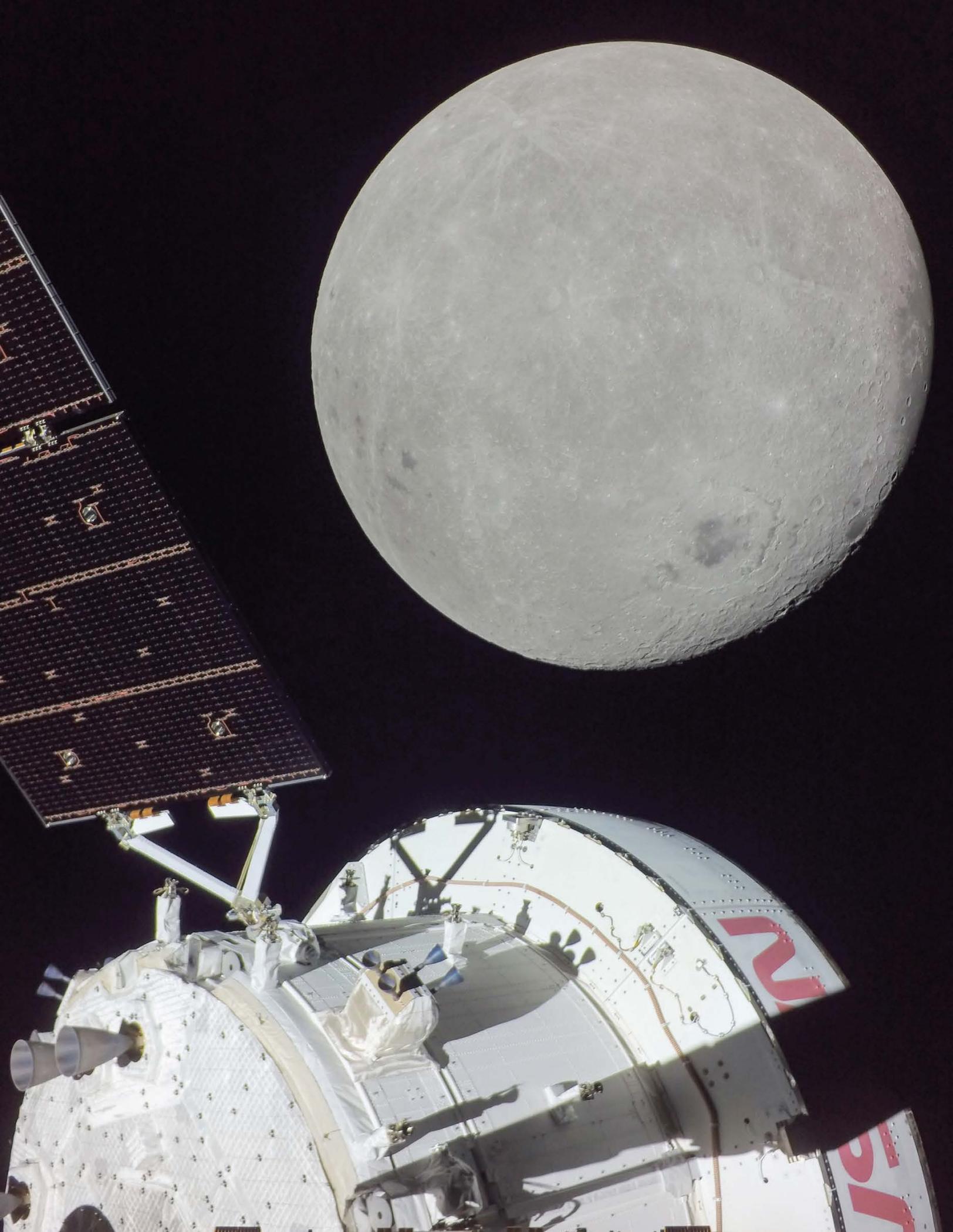